\documentclass[a4paper]{article}
\usepackage{ifthen}
\usepackage{epsfig}
\usepackage{amssymb}
\usepackage{graphicx}
\usepackage{amsmath}
\usepackage[center]{subfigure}
\renewcommand{\text}[1]{%
\ifthenelse{\equal{#1}{fB}}{f_B}{}%
\ifthenelse{\equal{#1}{fP}}{f_P}{}%
\ifthenelse{\equal{#1}{mB}}{m_B}{}%
\ifthenelse{\equal{#1}{xi0}}{\sigma_0}{}%
\ifthenelse{\equal{#1}{s0}}{s_0}{}%
\ifthenelse{\equal{#1}{M2}}{M^2}{}%
\ifthenelse{\equal{#1}{barxi}}{\bar\sigma}{}%
\ifthenelse{\equal{#1}{uu}}{u}{}%
\ifthenelse{\equal{#1}{ubar}}{\bar{\sigma}}{}%
\ifthenelse{\equal{#1}{vv}}{v}{}
\ifthenelse{\equal{#1}{phiBp}}{\phi^B_+}{}%
\ifthenelse{\equal{#1}{PhiBp}}{\phi^B_+}{}%
\ifthenelse{\equal{#1}{phiBmin}}{\phi^B_-}{}%
\ifthenelse{\equal{#1}{PhiBmin}}{\phi^B_-}{}%
\ifthenelse{\equal{#1}{PhiBpm}}{\overline\Phi^B_{\pm}}{}%
\ifthenelse{\equal{#1}{ppsiV}}{\Psi_V^B}{}%
\ifthenelse{\equal{#1}{ppsiA}}{(\Psi_A^B}{}%
\ifthenelse{\equal{#1}{bbarXA}}{\Psi_V^B}{}%
\ifthenelse{\equal{#1}{bbarYA}}{\overline Y_A^B}{}%
\ifthenelse{\equal{#1}{fV}}{f_V}{}%
\ifthenelse{\equal{#1}{mV}}{m_V}{}%
\ifthenelse{\equal{#1}{mP}}{m_P}{}%
\ifthenelse{\equal{#1}{lb}}{\lambda_B}{}%
}

\newcommand{\ba}{\begin{eqnarray}}
\newcommand{\ea}{\end{eqnarray}}
\newcommand{\be}{\begin{equation}}
\newcommand{\ee}{\end{equation}}
\newcommand{\DS}[1]{/\!\!\!#1}

\makeatletter
\def\fmslash{\@ifnextchar[{\fmsl@sh}{\fmsl@sh[0mu]}}
\def\fmsl@sh[#1]#2{%
  \mathchoice
    {\@fmsl@sh\displaystyle{#1}{#2}}%
    {\@fmsl@sh\textstyle{#1}{#2}}%
    {\@fmsl@sh\scriptstyle{#1}{#2}}%
    {\@fmsl@sh\scriptscriptstyle{#1}{#2}}}
\def\@fmsl@sh#1#2#3{\m@th\ooalign{$\hfil#1\mkern#2/\hfil$\crcr$#1#3$}}
\makeatother

\newcommand{\ps}[0]{\fmslash{p\,}\:\!}

\begin{document}
\begin{titlepage}
\begin{flushright}
SI-HEP-2006-03\\
\end{flushright}
\vfill
\begin{center}
{\Large\bf 
Form Factors from Light-Cone Sum Rules
with $B$-Meson Distribution Amplitudes}\\[2cm]
{\large\bf  
Alexander~Khodjamirian, Thomas Mannel and 
Nils Offen }\\[0.5cm]
{\it  Theoretische Physik 1, Fachbereich Physik,
Universit\"at Siegen,\\ D-57068 Siegen, Germany }\\
\end{center}
\vfill
\begin{abstract}
New sum rules 
for $B\to \pi,K $ and $B\to \rho,K^*$ form factors 
are derived from the correlation functions 
expanded near the light-cone in terms  of 
$B$-meson distribution amplitudes. 
The  contributions of quark-antiquark and quark-antiquark-gluon
components in the $B$ meson are taken into account. 
Models for the $B$-meson three-particle 
distribution amplitudes are suggested, based on QCD sum rules in HQET. 
Employing the new light-cone sum rules we calculate the form factors at 
small momentum transfers, 
including $SU(3)$-violation effects. 
The results agree with the predictions 
of the conventional light-cone sum rules.  
\end{abstract}
\vfill
\end{titlepage}

\section{Introduction}

There is a growing demand for more accurate and reliable 
calculations of heavy-to-light transition form factors in QCD.
The $B\to P,V$ form factors with $P=\pi,K$ and $V=\rho,K^*$
final states  provide the  hadronic input  
in exclusive semileptonic $B\to P(V) l \nu_l$, ~$B\to P(V) \bar{l}l$ 
and radiative  $B\to V \gamma $ decays. The same form factors
determine factorizable amplitudes in the nonleptonic charmless 
$B$-decays. All these decay channels are used for
determination of CKM parameters and for various tests 
of Standard Model. Pinning down the uncertainty  
of the form factors is in many cases the only way to 
increase the precision of these  analyses.

Lattice QCD is successfully used  to 
calculate heavy-to-light form factors
in the region of large momentum transfer squared,
$q^2=(p_B-p_{P,V})^2$.  To access  small $q^2$ 
(large energies of the light hadron), 
other QCD-based approaches are employed, such as  
the light-cone sum rules (LCSR) \cite{lcsr} for 
$B\to P$ \cite{BPLCSR,BallZw_BP} 
and $B\to V$ \cite{BVLCSR,BallZw_BV} form factors. 
To derive a LCSR, one starts with the operator-product
expansion (OPE) of a dedicated correlation function 
near the light-cone. The OPE result is then combined with the hadronic 
dispersion relation and quark-hadron duality, 
hence there are many common features 
with the original QCD sum rules \cite{SVZ}.
In the standard LCSR approach (hereafter called {\em light-meson LCSR}),
the correlation function is taken between the vacuum and light $P$- or 
$V$-meson state, whereas the $B$ meson is interpolated by a 
heavy-light quark current. 
As a result, the long-distance dynamics in the correlation
function is described by   
a set of pion, kaon, or $\rho$-, $K^*$-meson distribution amplitudes 
(DA's) of low twists. The main uncertainties
in the light-meson LCSR originate from the limited accuracy
of the  DA parameters. In addition, a sort of ``systematic'' uncertainty 
is brought by the quark-hadron duality approximation
in the $B$-meson channel. Hence, it is desirable 
to confirm the predictions of the light-meson LCSR 
by calculating the same form factors in an independent  way,
using different input and assumptions.

A different sum rule  for the $B\to \pi$ form factor 
was recently suggested by us in \cite{KMO} and, independently,
in the framework of SCET in \cite{DFH}.
The main idea is to  ``invert `` the correlation function,
that is, to interpolate the pion with an appropriate light-quark
(axial-vector) current, and put the $B$ meson on-shell
using the light-cone expansion in terms of the $B$-meson 
DA's. The latter are universal nonperturbative objects 
introduced in the framework of HQET \cite{GN} 
(see also \cite{SHB}; a review can be found in 
\cite{Grozin}) and used in 
several factorization  formulae for exclusive $B$-decays 
(see e.g., \cite{BF,blnugamma,BBNS,BVgamma}). 

In this paper the new version of LCSR
(we call it {\it $B$-meson LCSR}) is developed further. 
Following \cite{KMO}, we introduce 
the $B$-to-vacuum correlation function
and prove its light-cone dominance.  
The new  $B$-meson LCSR for several phenomenologically
important $B\to P,V$ form factors at $q^2\geq 0$ are derived. 
In addition to the leading-order contributions
of the two-particle (quark-antiquark) $B$-meson DA's 
$\phi_+^B$ and $\phi_-^B$, 
we calculate the corrections due to 
the three-particle (quark-antiquark-gluon) DA's defined in
\cite{Kawamura}. 
The functional form of the latter DA's was not known previously. 
Following \cite{GN,BIK}, we derive additional QCD sum rules 
for the vacuum correlation function of two heavy-light currents in HQET.    
We then use the perturbative parts of these sum rules to fix the behavior 
of the three-particle $B$-meson DA's at small light-cone momenta 
of the spectator quark and gluon. First models for three-particle
DA's are suggested in which the ``infrared'' behavior 
obtained from the HQET sum rules is combined
with the large-momentum fall-off. 
We find that the simple exponential ansatz 
for the two-particle DA's suggested in \cite{GN} 
and the exponential version of our model 
for the three-particle DA's form a selfconsistent set,
so that the relations between $B$-meson DA's   
\cite{Kawamura} following from the equations of motion
are fulfilled.

Our  model, as well as LCSR obtained below, 
do not include QCD radiative corrections which are 
beyond the scope of this work. 
NLO effects  have already been taken into account 
in more elaborated models of  $\phi_{+}^B$  
based on HQET sum rules \cite{BIK},
or on the first two moments \cite{LeeNeub}. 
The most important effect in NLO is 
the ``radiative tail'' of $\phi_{+}^B(\omega)$
at $\omega \to \infty$ caused by the nontrivial renormalization
\cite{NL} of the effective heavy-light current.  
Importantly, this peculiar ultraviolet behavior of 
$B$-meson DA's 
plays no role in LCSR at the leading, $O(\alpha_s^0)$ order,
where a sort of an end-point mechanism dominates. 
The duality threshold in the sum rule cuts
off the integration over the spectator momentum $\omega$ 
well below the 
region where the effect of the tail becomes noticeable.
Note that radiative corrections 
to the LCSR in SCET for the $B\to\pi$ form factor 
have already been calculated in \cite{DFH}, and their numerical
impact is moderate.

The LCSR obtained in this paper depend on the parameters 
determining the $B$-meson DA's. The most important input 
is  $\lambda_B$,  the first inverse moment  of  $\phi_+^B(\omega)$. 
At the same time, the new sum rules are  
independent of the DA's of $\pi,K$ or $\rho,K^*$ mesons. 
The light mesons are now  interpolated by the  
light-quark currents, hence the new sum rules  
rely on the quark-hadron duality in 
the channels of these currents.  The duality-threshold parameter 
in each channel is determined  from the corresponding 
two-point QCD (SVZ) sum rule for the light-meson decay constant. 
Furthermore, the $SU(3)$-violation effects are calculated 
in terms of the $s$-quark mass 
and the differences in the duality-threshold parameters for strange
and nonstrange mesons.  

From the new sum rules we obtain numerical predictions 
for various $B\to \pi,K$ and $B\to \rho,K^*$ form factors.
Our main observation is the sensitivity of the form factors
to the input value of $\lambda_B$. 
This circumstance was already used in \cite{KMO} to extract 
the interval for $\lambda_B$  using 
the light-meson LCSR result for the $B\to\pi$ form factor $f^+_{B\pi}$. 
Here, in order to be independent of the light-meson LCSR, we
use the interval of $\lambda_B$ inferred from QCD sum rules 
in HQET \cite{BIK}. With this input, we observe a good agreement of
the predicted form factors
with the most recent  results from the light-meson LCSR obtained 
in \cite{BallZw_BP,BallZw_BV}. 

In what follows, in section 2
we introduce the correlation function and discuss 
the applicability of the light-cone expansion.  
The sum rules in the leading order including 
the contributions of two- and three-particle DA's are derived 
in section 3. In section 4, the $B$-meson three-particle 
DA's are investigated and their form at small momenta 
of light-quark and gluon is established. The 
models of three-particle DA's are suggested 
and  the relations between two- and three-particle DA's
following from the equations of motion are investigated.  
In section 5 we discuss the heavy-mass dependence 
of the form factors obtained from $B$-meson LCSR.
Section 6 contains the numerical results for the form factors
and the concluding discussion.
In the Appendix  we present the bulky expressions
for the sum rules at nonvanishing light-quark mass and
nonzero momentum transfer.

\section{Correlation function}

Following \cite{KMO}, we define a generic correlation function 
of two quark currents 
sandwiched between the vacuum and the on-shell $\bar{B}$-meson state:  
\begin{equation}
F_{ab}^{(B)}(p,q)= i\int d^4x ~e^{i p\cdot x}
\langle 0|T\left\{\bar{q}_2(x)\Gamma_a q_1(x), 
\bar{q}_1(0)\Gamma_b b(0)\right\}|\bar{B}(P_B)\rangle\,,
\label{eq-corr}
\end{equation}
where $\bar{q}_1\Gamma_b b$ is one of the 
heavy--light (electro)weak 
currents and $\bar{q}_2\Gamma_a q_1$ is the interpolating current 
for a pseudoscalar or vector meson, with the flavour content 
determined by the valence quarks $q_{1,2}$.
The external momenta of the currents are  $q$ and $p$ respectively,
and $P_B^2=(p+q)^2=m_B^2$ . 
In Table 1 we list the combinations of quark flavours $q_1,q_2$ 
and Dirac-matrices $\Gamma_{a,b}$ for all $\bar{B}\to P,V$ transitions 
and  their form factors considered  in this paper. 
According to our choice, light pseudoscalar 
(vector) mesons are interpolated with the axial-vector 
(longitudinal vector) currents.

\begin{table}[t]
\begin{center}
\begin{tabular}{|c|c|c|c|c|c|}
\hline
Transition 
& $q_1$ & $q_2$ & $\Gamma_a$ & $\Gamma_b$ & Form factors \\
\hline\hline
&&&&&\\
& & & &$\gamma_\mu$ & $f_{B\pi}^+$, ~$f_{B\pi}^-$ \\
$\bar{B}\to \pi$ &$u$&$d,u$&$\gamma_\nu \gamma_5$ &&\\
& & & &$\sigma_{\mu\rho}$ & $f_{B\pi}^T$ \\
&&&&&\\
\hline
&&&&&\\
& & & 
&$\gamma_\mu$ & $f_{BK}^+$, $f_{BK}^-$ \\
$\bar{B}\to K $ & $s$ & $d,u$ &$\gamma_\nu\,\gamma_5$ & &\\
&&
& &$\sigma_{\mu\rho}$ & $f_{BK}^T$ \\
&&&&&\\
\hline
&&&&&\\
& &  & & $\gamma_\mu$  & $V^{B\rho}$ \\
&&&&&\\
$\bar{B}\to \rho$ & $u$ & $d.u$  &$\gamma_\nu$ 
&$\gamma_\mu \gamma_5$ &$A^{B\rho}_1$, $A^{B\rho}_2$\\ 
&&&&&\\
& & & &$\sigma_{\mu\rho}$  & $T^{B\rho}_1$ \\
&&&&&\\
\hline
&&&&&\\
& & & & $\gamma_\mu$ & $V^{BK^*}$ \\
&&&&&\\
$\bar{B}\to K^*$ & $s$ & $d,u$ & $\gamma_\nu$ &$\gamma_\mu\gamma_5$ &
$A^{B K^*}_1$,~ $A^{BK^*}_2$\\
&&&&&\\
& & & & $\sigma_{\mu\rho}$  & $T^{B K^*}_1$ \\
&&&&&\\
\hline
\end{tabular}
\end{center}
\caption{\it Combinations of light-quark flavours and Dirac matrices
in the correlation function (\ref{eq-corr}) 
and the corresponding heavy-to-light 
form factors.}
\label{tab-list}
\end{table}

First of all, we have to  
convince ourselves that OPE 
near the light-cone is applicable for the correlation 
function (\ref{eq-corr}) if the variables  
$p^2$ and $q^2$ are far below the 
hadronic thresholds in the channels of
$\bar{q_2}\Gamma_a q_1$ and $\bar{q_1}\Gamma_b b$ 
currents, respectively.  
The correlation function  can be systematically expanded in 
the limit of large $m_b$ in HQET. 
Separating the static momentum of the $B$-meson state, 
we rewrite $P_B=p+q=m_bv +k$, 
where $v$ is the four-velocity of $B$, and $k$ is the residual 
momentum. We retain the relativistic normalization of the state: 
$|B(P_B)\rangle=|B_v\rangle$, up to $1/m_b$ corrections . 
Also the $b$-quark field 
is substituted by the effective field, using $b(x) = e^{-im_bvx} h_v(x)$. 
For simplicity we consider the rest frame $v=(1,0,0,0)$. 
In first 
approximation, $m_B= m_b+\bar \Lambda$, hence 
$k_0\sim \bar{\Lambda}$ in this frame.  
We also redefine  
the four-momentum transfer $q$ by separating the ``static'' part of it: 
$q=m_b v+\tilde{q}$, so that $p+\tilde{q}=k$.
After the transition to HQET,
\begin{equation}
F_{ab}^{(B)}(p,q)=\tilde{F}_{ab}^{(B_v)}(p,\tilde{q})+O(1/m_b)\,,
\label{eq-HQET}
\end{equation}
the correlation function in the heavy $m_b$ limit,
\begin{equation}
\tilde{F}_{ab}^{(B_v)}(p,\tilde{q})= i\int d^4x ~e^{i p\cdot x}
\langle 0|T\left\{\bar{q}_2(x)\Gamma_a q_1(x), 
\bar{q}_1(0)\Gamma_b h_v(0)\right\}|\bar{B}_v\rangle 
\label{eq-HQETcorr}
\end{equation}
does not depend on $m_b$, if $p^2$ and $\tilde{q}^2$ are 
generic scales. In this amplitude 
two light-quark currents (one of them containing the effective 
field $h_v$) with virtualities $p^2$ and $\tilde{q}^2$  
annihilate an effective hadronic state 
with a mass of $O(\bar{\Lambda})$. From the QCD point of view,
the correlation function (\ref{eq-HQETcorr}) resembles  
the $\gamma^*(p)\gamma^*(\tilde{q})\to \pi^0(p+\tilde{q})$ 
transition amplitude.
For the latter a detailed proof of the light-cone dominance can be found, 
e.g.,  in \cite{CK}. Following the same line of arguments 
for the amplitude  $\tilde{F}_{ab}^{(B_v)}(p,\tilde{q})$,     
we assume that both four-momenta are  spacelike, 
$p^2,\tilde q^2<0$, and  sufficiently large: 
\be 
P^2 , |\tilde{q}^2|\gg \Lambda_{QCD}^2,
\bar{\Lambda}^2\,,
\label{eq-scale2}
\ee
where $P^2\equiv -p^2$. Simultaneously, the difference 
between the virtualities is kept large, so that the ratio 
\be  
\xi=\frac{2 p\cdot k }{P^2}\sim \frac{|\tilde{q}^2|-P^2}{P^2}
\neq 0,\  
\label{eq-xi}
\ee
is at least of $O(1)$.
With these conditions fulfilled,
the integral in (\ref{eq-HQETcorr}) 
is supported in the region of small $x^2\leq 1/P^2$,
where the exponent  $e^{ipx}$ does not oscillate strongly. 

Returning to the momentum-transfer squared   $q^2$, 
one obtains 
\ba
q^2\simeq m_b^2+2m_b\tilde{q}_0~\sim  m_b^2 -m_b P^2\xi/\bar{\Lambda}\,.
\label{eq-tau}
\ea
Thus, $q^2$ is far from the threshold 
$\sim m_b^2$  in the heavy-light channel, 
if the conditions (\ref{eq-scale2}) and (\ref{eq-xi}) 
are fulfilled. Parametrically,  
the lower part of the physical region of  $B\to P,V$ transitions 
\be
0\leq q^2 <  m_b^2 -m_b P^2/\bar{\Lambda} 
\label{eq-interv}
\ee
is accessible to OPE on the light-cone. 
One encounters a situation 
similar to the light-meson LCSR which are applicable  
up to $q^2=m_b^2-m_b\chi$,  where $\chi = O( 1 \mbox{GeV})$ 
does not scale with $m_b\to\infty$.
For example, the LCSR with pion DA's \cite{BPLCSR} 
can be used up to  $q^2\leq 14-16$ GeV$^2$.    
For the $B$-meson LCSR considered here, the upper limit
of the interval (\ref{eq-interv}) could not be that large, 
because  generally $P^2/\bar{\Lambda}\gg \chi$. 
An important case is when $q^2$ is in the vicinity of zero.
Solving Eq.~(\ref{eq-tau})  for $q^2=0$ one obtains  
$P^2\xi\sim m_b\bar{\Lambda}$. With $P^2$ being large but independent
of $m_b$, in this case $\tilde{q}^2$ scales with $m_b\to\infty$:
$|\tilde{q}^2|=P^2(1+\xi) \sim m_b\bar{\Lambda}$.

The light-cone dominance of the 
correlation function allows one to  
contract the $q_1$ and $\bar{q}_1$ fields and use the free-quark propagator 
$S_{q_1}(x)=-i\langle 0 | q_{1}(x)\bar{q}_{1}(0)|0\rangle$ 
as a leading-order approximation. 
The corresponding diagram is depicted in Fig.~1a.
We obtain from  Eq.~(\ref{eq-HQETcorr}) 
(neglecting for simplicity the light-quark mass $m_{q_1}$):   
\begin{equation}
\tilde{F}_{ab}^{(B_v)}(p,\tilde{q})= i\int d^4x ~e^{i p\cdot x}
\frac {ix_\rho}{2\pi^2(x^2)^2}[\Gamma_a \gamma_\rho \Gamma_b]_{\alpha\beta} 
\langle 0|\bar{q}_{2\alpha}(x) h_{v\beta}(0)|\bar{B}_v\rangle\,, 
\label{eq-leadcorr}
\end{equation}
a convolution of a short-distance part with the matrix element 
of the bilocal operator between the vacuum and $B_v$-state. Expanding the operator 
$\bar{q}_{2\alpha}(x) h_{v\beta}(0)$ 
at $x=0$ one encounters, in the generic case $\xi\sim 1$, 
an infinite  series of matrix elements of local operators, 
as  explained in details in \cite{BKM,CK} for the  
vacuum-pion amplitudes. Instead, one has to retain in Eq.~(\ref{eq-leadcorr})
the matrix element of the bilocal operator, expanding it around $x^2=0$. 
This procedure brings the $B$-meson DA's 
into the game. They however do not have a well defined twist, 
contrary to the light-meson DA's. 
The definitions of two- and three-particle  $B$-meson
DA's at the leading order of $x^2\to 0$ expansion 
will be given in the next section. 
\begin{figure}[t]
\includegraphics[width=10cm]{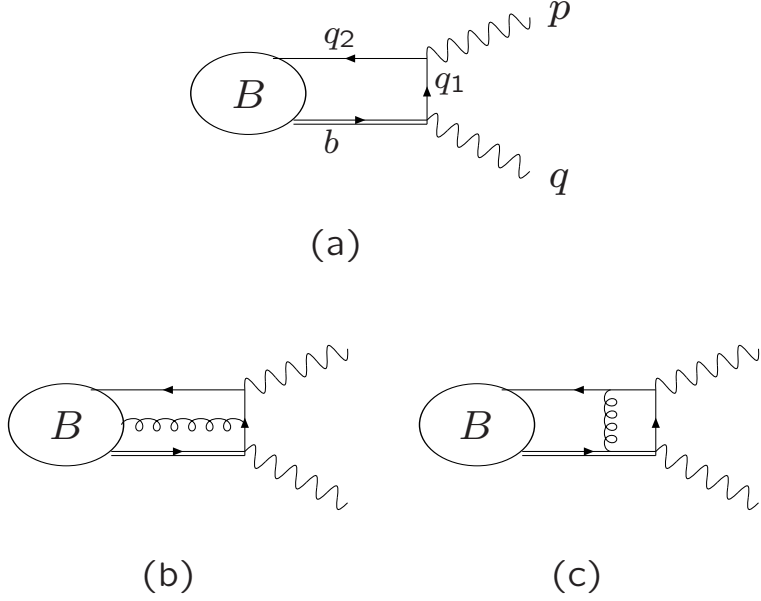}\\
\caption{ \it Diagrams corresponding to 
the contributions of 
(a) two-particle and (b) three-particle $B$ -meson DA's 
to the correlation function
(\ref{eq-corr}); (c)- one of the $O(\alpha_s)$ diagrams.
Curly (wavy) lines denote gluons (external currents).}
\label{fig-diags}
\end{figure}

\section{Derivation of LCSR}

The sum rules are obtained following the standard procedure \cite{SVZ},
that is, matching the OPE result for the correlation 
function to the hadronic representation and employing 
quark-hadron duality and Borel transformation. 
Considering, for definiteness, the case when 
the current $\bar{q}_2\Gamma_a q_1$ 
interpolates a pseudoscalar meson $P$ ($\pi$ or $K$),
we write  the correlation function 
(\ref{eq-corr}) in a form of  the hadronic 
dispersion relation in the channel of the light meson:
\be
F_{ab}^{(B)}(p,q) =\frac{\langle 0|\bar{q}_2\Gamma_a q_1|P(p)\rangle\langle
  P(p)|\bar{q}_1\Gamma_b b|\bar{B}(P_B)\rangle
}{m_{P}^2-p^2}+...\,,
\label{eq-disp}
\ee
where only the $P$-meson pole term
is shown explicitly, and ellipses indicate 
the contributions of excited and continuum states. The two hadronic matrix 
elements in Eq.~(\ref{eq-disp}) are determined, respectively,  
by the decay constant of $P$ and by the $B\to P$ form factors.

To proceed, the dispersion relation (\ref{eq-disp})
is equated to the HQET correlation function (\ref{eq-leadcorr}), 
which will be calculated using light-cone OPE:  
\be
F_{ab}^{(B),OPE}(p,q)\simeq \tilde{F}_{ab}^{(B_v),OPE}(p,q-m_bv).
\label{FOPE}
\ee
After Lorentz-decomposition,
$$F_{ab}^{(B),OPE}(p,q)= l_{ab}(p,q)F^{(B),OPE}(p^2,q^2)+...\,,$$
each invariant amplitude is conveniently represented in a 
form of dispersion relation:
\be
F^{(B),OPE}(p^2,q^2) =\frac1\pi\int\limits_{(m_{q_1}+m_{q_2})^2}^\infty ds\,
\frac{\mbox{Im}F^{(B),OPE}(s,q^2)}{s-p^2}\,,
\label{eq-dispOPE}
\ee
where the lower threshold is given by the sum of the light-quark masses.
Furthermore, employing quark-hadron duality 
approximation we equate the P-meson contribution 
in Eq.~(\ref{eq-disp})  to the part of the dispersion
integral (\ref{eq-dispOPE}) limited from above by  the effective threshold 
$s_0^P$. After Borel transformation, 
the LCSR for the $B\to P$ matrix element can be written 
down in the following generic form:
\ba
\langle 0|\bar{q}_2\Gamma_a q_1|P(p)\rangle\langle
  P(p)|\bar{q}_1\Gamma_b b|\bar{B}(P_B)\rangle
e^{-m_P^2/M^2}
\nonumber\\
=l_{ab}(p,q)\int\limits_{(m_{q_1}+m_{q_2})^2}^{s_0^{P}}ds\,e^{-s/M^2}
\mbox{Im} F^{(B),OPE}(s,q^2)+...\,,
\label{eq-LCSR}
\ea
where the ellipses denote the rest of the Lorentz-decomposition.
The derivation of LCSR in the case of a vector meson $V$ ($\rho$ or $K^*$) is 
fully analogous, with $V$ replacing $P$ in Eqs.~(\ref{eq-disp}),
(\ref{eq-LCSR}). 
For each combination of currents listed
in Table~1, using the definitions of the hadronic 
matrix elements given below and decomposing 
the l.h.s of Eq.~(\ref{eq-LCSR}) in  
invariant amplitudes 
it is straightforward to 
obtain a separate sum rule for a given form factor.

The following standard definitions are used 
for the decay constants
of pseudoscalar and vector mesons:
\ba
&&\kappa \langle 0|\bar{q}_2\gamma_\nu\gamma_5 q_1|P(p)\rangle
=i p_\nu f_P  ,\nonumber \\
&&\kappa \langle 0|\bar{q}_2\gamma_\nu q_1|V(p)\rangle=
\epsilon^V_\nu m_Vf_V\,,
\label{decconst}
\ea
for $B\to P$ form factors:  
\begin{eqnarray}
&&\kappa\langle P(p)|\bar{q_1}\gamma_\mu b| \bar{B}(p+q)\rangle
=2p_\mu f^+_{BP}(q^2)+q_\mu \left[
f^+_{BP}(q^2)+f^-_{BP}(q^2)\right]\,,
\nonumber
\\
&&\kappa\langle P(p)|\bar{q_1}\sigma_{\mu\rho}q^\rho b| B(p+q)\rangle
=\left[q^2(2p_\mu+q_\mu)-(m_B^2-m_P^2)q_\mu\right]\dfrac{i\,f_{BP}^T(q^2)}{m_B+m_P}\,,\nonumber\\
\label{eq-formfBP}
\end{eqnarray} 
and for $B\to V$ form factors:
\begin{eqnarray}
\kappa\langle V(p)|\bar{q_1}\gamma_\mu(1-\gamma_5) b| \bar{B}(p+q)\rangle
=-i\epsilon_\mu^*(m_B+m_V)A_1^{BV}(q^2)\nonumber\\
+i(2p+q)_\mu(\epsilon^*q)\dfrac{A_2^{BV}(q^2)}{m_B+m_V}+iq_\mu(\epsilon^* q)\dfrac{2m_V}{q^2}\Big(A_3^{BV}(q^2)-A_0^{BV}(q^2)\Big)\nonumber\\
+\epsilon_{\mu\nu\rho\sigma} 
\epsilon^{*\nu}q^\rho p^\sigma\dfrac{2 V^{BV}(q^2)}{m_B+m_V}\,,
\label{eq-formfBV}
\end{eqnarray}
with $2m_V A_3^{BV}(q^2)= (m_B+m_V)A_1(q^2)-(m_B-m_V)A_2^{BV}(q^2)$ and
$A_0^{BV}(0)=A_3^{BV}(0)$, and 
\begin{eqnarray}
\kappa\langle V(p)|\bar{q_1}\sigma_{\mu\rho}q^\rho(1+\gamma_5) b| \bar{B}(p+q)\rangle
=i\epsilon_{\mu\nu\rho\sigma}\epsilon^{*\nu}q^\rho p^\sigma\,2\,T_1^{BV}(q^2)\nonumber\\
+\{\epsilon^*_\mu(m_B^2-m_V^2)-(\epsilon^* q)(2 p+q)_\mu\}T_2^{BV}(q^2) 
\nonumber\\
+(\epsilon^*q)\left\{q_\mu-\dfrac{q^2}{m_B^2-m_V^2}(2p+q)_\mu\right\}T_3^{BV}(q^2)\,.
\label{eq-formfBVT}
\end{eqnarray}
In the above, $\kappa=\sqrt{2} ~(\kappa=1)$ 
for $\pi^0$ and $\rho^0$ (for other mesons). 

In what follows, we 
derive new  LCSR for the  $B\to P,V$ form factors 
listed in Table 1. For definiteness, we 
assume the following flavour configurations: $\bar{B}_d^0\to\pi^+,\rho^+$
and $\bar{B}_d^0\to \bar{K}^0,\bar{K}^{*0}$.
The sum rules for the remaining form factors  
$f^0_{BP}$, $A_0^{BV}$ and $T_{2,3}^{BV}$,  will be presented elsewhere. 
Importantly, in all channels
considered in this paper, the threshold parameters $s_0^{P,V}$ 
can be obtained from the two-point sum rules for 
the decay constants  $f_{P,V}$.    

Let us now calculate the r.h.s. of the sum rule
(\ref{eq-LCSR}). As explained in the previous section, 
the leading-order contribution to the OPE is  
given by the diagram in Fig.~1a. The answer is obtained, 
by decomposing the matrix element in Eq.~(\ref{eq-leadcorr}) 
at $x^2=0$:  
\begin{eqnarray}
&&
\langle 0|\bar{q}_{2\alpha}(x)[x,0] h_{v\beta}(0)
|\bar{B}_v\rangle
\nonumber \\ 
&&
= -\frac{if_B m_B}{4}\int\limits _0^\infty 
d\omega e^{-i\omega v\cdot x} 
\left [(1 +\DS v)
\left \{ \phi^B_+(\omega) -
\frac{\phi_+^B(\omega) -\phi_-^B(\omega)}{2 v\cdot x}\DS x \right \}\gamma_5\right]_{\beta\alpha}
\label{eq-BDAdef}
\end{eqnarray}
in terms of  the $B$-meson two-particle DA's 
$\phi_{+}^B(\omega)$ and $\phi_{-}^B(\omega)$ defined 
\cite{GN,BF} in the momentum space.  
In the above, ~$[x,0]$ is the  path-ordered gauge factor.  
The variable $\omega>0$ is the plus component of the 
spectator-quark momentum in the $B$ meson. 
Substituting Eq.~(\ref{eq-BDAdef}) in Eq.~(\ref{eq-leadcorr})
and integrating over $x$, one obtains the  
invariant amplitudes $F^{(B),OPE}(p^2,q^2)$
which have a simple generic expression at $q^2=0$:
\be
F^{(B),OPE}(p^2,0)=\sum\limits_{n=1,2}
\int\limits_0^\infty \frac{d\omega \,\phi_n(\omega)}
{\left[(1-\omega/m_B)(\omega m_B-p^2)\right]^n}\,,
\label{eq-inv}
\ee
where the functions $\phi_n(\omega)$ 
are combined from the $B$-meson DA's.
If one continues the $x^2$-expansion 
of the matrix element  (\ref{eq-BDAdef}) beyond the 
leading order, the resulting contributions 
to $F^{(B),OPE}$ will be suppressed by 
additional powers of the denominator (i.e., by inverse powers
of $M^2$ after Borel transformation). We neglect
them, having assumed that $P^2=-p^2$ (or $M^2$) 
is a large scale. Furthermore, $B$-meson DA's are essentially 
concentrated  around $\omega\sim \bar{\Lambda}$, 
where $\bar{\Lambda}=m_B-m_b$, with the kinematical limit 
$\omega <2 \bar{\Lambda}$. Hence, the denominator 
in Eq.~(\ref{eq-inv}) implicitly contains another 
large  scale $m_B\bar{\Lambda}$. 
The heavy-mass scale which reappears in the HQET correlation function,
has a kinematical origin: at $q^2=0$ the external momenta
$p$ and $q$ are both $O(m_b/2)$, or in other words, as already 
mentioned in the previous section, the rescaled virtuality 
$\tilde{q}^2=O(m_b\bar{\Lambda})$.
Finally, to obtain the r.h.s. of Eq.~(\ref{eq-LCSR}), 
one transforms the integral in Eq.~(\ref{eq-inv})
into a dispersion form, changing the variable 
$\omega$ to $s=\omega m_B$, performing the Borel transformation 
in the variable $p^2$ and replacing the upper limit by the duality 
threshold $s_0^{P,V}$.
Importantly, due to the fact that  
$\sqrt{s_0^{P,V}}\ll m_B$ 
only the regions of small momenta of spectator quark 
$\omega < s_0^{P,V}/m_B$, far from the kinematical
threshold $\omega \sim \bar{\Lambda}$ are important
in the LCSR. 
As already mentioned in \cite{KMO}, this situation 
corresponds to the end-point mechanism which is realized 
in heavy-to-light exclusive transitions in the absence of 
hard-gluon exchanges.  

Following the derivation described above, 
we obtain the leading-order 
LCSR for the $B\to \pi,\rho$ form factors 
at zero momentum transfer ($q^2=0$), where the $u,d$-quark 
masses are neglected, and the pion mass is put to zero:
\ba
f^+_{B\pi}(0)=
\frac{f_B }{ f_\pi\,m_B}
    \int\limits _0^{s_0^\pi} ds e^{-s/M^2}
    \text{PhiBmin}(s/m_B)\,,  
\label{eq-fplBpi0}
\ea

\ba
f^+_{B\pi}(0)+f^-_{B\pi}(0)
=\dfrac{f_B}{f_\pi\,m_B}\int_0^{s_0^\pi}ds e^{-s/M^2}\left[\dfrac{m_B^2}{m_B^2-s}\text{PhiBp}(s/m_B )  
\right.
\nonumber \\
\left.
-\,2\dfrac{s}{m_B^2-s}\,\text{PhiBmin}(s/m_B )\,+\,2\dfrac{m_B^3}{(m_B^2-s)^2}\text{PhiBpm}(s/m_B )\right]\,,
\label{eq-fpmBpi0}
\ea

\ba
f^T_{B\pi}(0) = \dfrac{f_B}{f_\pi\,m_B}\int_0^{s_0^\pi} ds e^{-s/M^2}\left[\text{PhiBmin}(s/m_B )\right.\nonumber\\
\left.
-\,\text{PhiBp}(s/m_B )\,-\,\dfrac{m_B}{m_B^2-s}\,\text{PhiBpm}(s/m_B)\right]\,,
\label{eq-fTBpi0}
\ea

\ba
V^{B\rho}(0)= \dfrac{f_B (m_B\,+\,m_\rho)}{2\,f_\rho\,m_\rho m_B}
\,e^{m_\rho^2/M^2}\int_0^{s_0^\rho} ds e^{-s/M^2}
\dfrac{m_B^2}{m_B^2-s}\text{PhiBp}(s/m_B)\,,
\label{eq-VBrho0}
\ea

\ba
A_1^{B\rho}(0)=\dfrac{f_B m_B }{2\,f_\rho\,m_\rho (m_B+m_\rho)}\,e^{m_\rho^2/M^2}\int_0^{s_0^\rho} ds e^{-s/M^2}\text{PhiBp}(s/m_B)\,,
\label{eq-A1Brho0}
\ea

\begin{multline}
A_2^{B\rho}(0)=\dfrac{f_B}{2\,f_\rho\,m_\rho}\,\dfrac{(m_B\,+\,m_\rho)}{m_B}\,e^{m_\rho^2/M^2}\int_0^{s_0^\rho} ds e^{-s/M^2}\\
\left[\dfrac{m_B^2}{m_B^2-s}\text{PhiBp}(s/m_B)
-\,2\dfrac{s}{m_B^2-s}\text{PhiBmin}(s/m_B )\,+\,2\dfrac{m_B^3}{(m_B^2-s)^2}\text{PhiBpm}(s/m_B)\right]\,,
\label{eq-A2Brho0}
\end{multline}

\ba
T_1^{B\rho}(0)= \dfrac{f_B}{2\,f_\rho\,m_\rho}\,e^{m_\rho^2/M^2}\int_0^{s_0^\rho} ds e^{-s/M^2} \text{PhiBp}(s/m_B)   
\label{eq-T1Brho0}\,,
\ea

where a compact notation: 
\be
\text{PhiBpm}(\omega )=\int\limits_0^{\omega} d\tau \Big(\phi^B_+(\tau )-\phi^B_-(\tau)\Big)
\nonumber
\ee 
is introduced.
The Borel parameter $M$  in the light-meson channels 
has typical values around $1$ GeV, still  $M\gg\Lambda_{QCD}$.
The first sum rule (\ref{eq-fplBpi0}) has already been derived 
in \cite{KMO} (see also \cite{DFH}),
whereas all other sum rules are new. 
The LCSR at $q^2\neq 0$ and 
$m_{q_1}\neq 0 $ have bulky expressions 
presented in the Appendix. Substituting $m_{q_1}=m_s$
and replacing $s_0^{\pi,\rho}\to s_0^{K,K^*}$, one 
obtains LCSR  for the $B\to K,K^*$ form factors.

In this paper we neglect $O(\alpha_s)$ 
radiative corrections 
due to the hard-gluon exchanges between the quark-antiquark lines
(one of the diagrams is shown in Fig. 1c). Their calculation is 
inseparable from the nontrivial renormalization of $B$-meson DA's,
which is so far known only for $\phi_{+}^B(\omega)$ \cite{NL} 
(for a detailed discussion see also \cite{BIK,Grozin}). As far as 
the normalization
scale of $\phi_{+}^B(\omega)$ or its inverse moment 
\be
\frac{1}{\lambda_B(\mu)}=\int_0^\infty d\omega \frac{\phi_{+}^B(\omega,\mu)}{\omega}
\label{omega}
\ee
is concerned, we assume that $\mu\simeq M$, having in mind that 
the Borel scale reflects the average
virtuality in the correlator.

In addition we calculate the corrections due 
to the three-particle (quark-antiquark-gluon) DA's
of the $B$ meson. They correspond to the diagram in Fig.~1b,
where a low virtuality gluon is emitted from the virtual quark
and absorbed in the $B$ meson. The  
contribution of this diagram to the correlation function (\ref{eq-corr})
is obtained by  contracting the $q_1(x)$ and $\bar{q}_1(0)$ fields 
and inserting the one-gluon part of the 
quark propagator near the light-cone \cite{BB}:
\ba
S_{q_1}(x,0,m_{q_1})=\!\int\!\dfrac{d^4p}{(2\pi)^4}e^{-ipx}
\!\int_0^1\!dv\,G_{\mu\nu}(vx)
\left[\dfrac{v x^\mu
    \gamma^\nu\,}{p^2-m_{q_1}^2}-\dfrac{(\ps+m_{q_1})
\sigma^{\mu\nu}}{2(p^2-m_{q_1}^2)^2}
\right]\,,
\ea
where $G_{\mu\nu}=g_sG^a_{\mu\nu}(\lambda^a/2)$.  
As a result, an expression
similar to Eq.~(\ref{eq-leadcorr})
emerges with the vacuum-to-$B$ matrix element containing a nonlocal 
product of quark, antiquark and gluon fields. 
In $x^2=0$ limit we adopt the following 
decomposition of  this matrix element into 
four independent  three-particle DA's: 
\begin{eqnarray}
&&\langle 0|\bar{q_2}_\alpha(x) G_{\lambda\rho}(ux) 
h_{v\beta}(0)|\bar{B}^0(v)\rangle=
\frac{f_Bm_B}{4}\int\limits_0^\infty d\omega
\int\limits_0^\infty d\xi\,  e^{-i(\omega+u\xi) v\cdot x} 
\nonumber \\ 
&&\times \Bigg [(1 +\DS v) \Bigg \{ (v_\lambda\gamma_\rho-v_\rho\gamma_\lambda)
\Big(\Psi_A(\omega,\xi)-\Psi_V(\omega,\xi)\Big)
-i\sigma_{\lambda\rho}\Psi_V(\omega,\xi)
\nonumber\\
&&-\left(\frac{x_\lambda v_\rho-x_\rho v_\lambda}{v\cdot x}\right)X_A(\omega,\xi)
+\left(\frac{x_\lambda \gamma_\rho-x_\rho \gamma_\lambda}{v\cdot x}\right)Y_A(\omega,\xi)\Bigg\}\gamma_5\Bigg]_{\beta\alpha}\,,
\label{eq-B3DAdef}
\end{eqnarray}
where the path-ordered gauge factors on l.h.s. are omitted for
brevity. Multiplying both parts of this expression by 
$x^\rho$ one encounters the definition  
introduced in  \cite{Kawamura}. 
The DA's $\Psi_{V}$,$\Psi_{A}$, $X_A$ and $Y_A$ depend 
 on the two variables $\omega>0$ and $\xi>0$ being, respectively, the plus components of the light-quark and gluon momenta 
in the $B$ meson.

Our analysis in this paper is restricted to the four 
three-particle DA's defined in Eq.~(\ref{eq-B3DAdef}), 
and to the  two-particle  DA's defined 
in Eq.~(\ref{eq-BDAdef}). One can further 
expand both matrix elements near the light-cone in powers of 
$x^2$  introducing additional DA's \footnote{Recently, a more 
general decomposition of the three-particle 
matrix element (\ref{eq-B3DAdef}) was suggested  
in \cite{GW}, where one encounters 
additional three-particle DA's. A separate 
study is needed to clarify the importance 
of these amplitudes with respect to the main four 
three-particle DA's.}. As argued above, their
contributions to the correlation function  will be 
power-suppressed, at least by inverse powers of $M^2$.

The resulting expressions for the three-particle 
contributions to LCSR at $q^2\neq 0$
are presented in the Appendix. At $q^2=0$  these expressions 
(which we do not display for brevity) have to be added 
to the leading-order sum rules (\ref{eq-fplBpi0})-(\ref{eq-T1Brho0}). The new LCSR are sensitive to the normalization 
constants and to the behavior of the B-meson two-particle 
(three-particle) DA's  at small $\omega$ ($\omega,\xi$),
hence, also to the inverse moment $\lambda_B$.
For the two-particle DA's the behavior at $\omega\to 0$ 
is known, 
and we have at our disposal models for $\phi^B_\pm(\omega)$ 
\cite{GN,BIK,LeeNeub}. The remaining task is to establish 
the behavior of the three-particle DA's at small $\omega,\xi$ , 
and to build a model for them. 

\section{Three-particle DA's from sum rules in HQET}

As already mentioned, QCD sum rules in HQET were employed in \cite{GN,BIK} to 
predict the $B$-meson two-particle DA's $\phi_{\pm}^B(\omega)$. 
The idea was to introduce a correlation 
function with two  $\bar{q} \Gamma h_v$ currents, one of them 
local and the other one containing  
the $h_v$ and $\bar{q}$ fields at a 
light-like separation. The ground $B_v$-state contribution 
to the hadronic dispersion relation for this correlation function   
contains the product of the $B$-meson decay constant and the nonlocal 
heavy-to-light matrix element (\ref{eq-BDAdef}). An appropriate choice of 
the Dirac-structure $\Gamma$ allows one to separate $\phi_{+}^B$ from 
$\phi_{-}^B$. Matching the $B$-meson term to the leading perturbative 
contribution (the loop diagram) via quark-hadron duality, 
one reproduces \cite{GN} the behavior of both DA's at $\omega\to 0$: 
$\phi_{+}^B(\omega)\sim \omega$ and  $\phi_{-}^B(\omega)\sim const$, 
in accordance with general expectations.

Here we use a similar method 
and derive HQET sum rules for the $B$ meson three-particle DA's in 
the perturbative loop approximation. 
\begin{figure}
\begin{center}
\epsfig{file=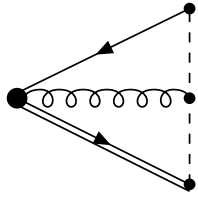,scale=1.3}
\end{center}
\caption{\it Perturbative loop diagram for 
the correlation function (\ref{eq-corrHQET}). 
The points connected with the dashed line
(the thick point)   
represent the vertex of the nonlocal (local) current.}
\end{figure}
The starting object is the correlation function 
\begin{multline}
\Pi^{(\Gamma)}_\lambda(\rho,t,u)\:=\:i\int d^4y\,e^{-i\rho(v\cdot y)}\\
\langle0\vert T\{\bar{q}(tn)\Gamma
G_{\lambda\sigma}(utn)n^\sigma h_v(0),\bar{h}_v(y)
G_{\alpha\beta}\sigma^{\alpha\beta}\gamma_5q(y)\}\vert0\rangle\,,
\label{eq-corrHQET}
\end{multline}
where the local current containing the effective heavy-quark, 
light-antiquark and gluon fields is correlated with a 
generic nonlocal current, with all three fields on the light-cone. We define the light-like unit vectors $n_\mu$  and $\bar{n}_\mu$ 
($n^2=\bar{n}^2=0,n\cdot\bar{n}=2$)
so that $v_\mu=(n_\mu+\bar{n}_\mu)/2$; $t$ is an arbitrary real number,
determining the location on the light-cone (that is, $tn$ corresponds 
to the light-like interval $x$ in 
Eq.~(\ref{eq-B3DAdef})) and $\rho$ is the 
``off-shell energy``, the HQET analog of virtuality. The 
 gauge factors between the fields in Eq.~(\ref{eq-corrHQET}) 
are omitted for brevity, in fact they are
inessential for the perturbative loop approximation.  
The local current in (\ref{eq-corrHQET}) is chosen in a convenient 
scalar form, note that other choices are also possible. 

The correlation function (\ref{eq-corrHQET}), after 
inserting the complete set of hadronic states, has a pole 
of the $B_v$ state at $\rho=\bar{\Lambda}$ where 
$\bar{\Lambda}=m_B-m_b$,
schematically:
\begin{equation}
\Pi^{(\Gamma)}_\lambda(\rho,t,u)=
\dfrac{C_\lambda}{\bar{\Lambda}\,-\,\rho
}\int_0^\infty d\omega
\int_0^\infty d\xi\,e^{-i(\omega+u\xi)t} F(\omega,\xi)\,,
\label{eq-HQETpole}
\end{equation}
where $C_\lambda$ is  proportional to the hadronic 
matrix element of the local current in Eq.~(\ref{eq-corrHQET})
and to other normalization constants; 
since we are only interested in the functional 
dependence on $\omega$ and $\xi$,
this factor does not need to be specified. 
$F(\omega,\xi)$  is one of the three-particle DA's 
$\Psi_A(\omega,\xi),\,\Psi_V(\omega,\xi),\,X_A(\omega,\xi)$,
$Y_A(\omega,\xi)$ (or their linear combination),  depending on the choice
of the Dirac-structure in the nonlocal current 
in Eq.~(\ref{eq-corrHQET}). More specifically, the following
correspondence is established:  
\begin{eqnarray}
\Gamma=\gamma_\mu\gamma_5&\rightarrow &
F=\Psi_A\nonumber\\
\gamma_\mu&\rightarrow&\;\;\;\;\;\;\;\;\Psi_V\nonumber\\
\sigma_{\mu\nu}\gamma_5&\rightarrow&\;\;\;\;\;\;\;\;\Psi_A-\Psi_V\nonumber\\
\fmslash{n}\gamma_5&\rightarrow&\;\;\;\;\;\;\;\;X_A+\Psi_A\nonumber\\
i\gamma_5 & \rightarrow &\;\;\;\;\;\;\;\; Y_A\,-\,X_A\,.
\label{eq-3loop}
\end{eqnarray}

To proceed, we calculate the spectral density of the 
leading-order perturbative contribution to the 
correlation function (\ref{eq-corrHQET}), given by 
the loop diagram in Fig.~2.
All three intermediate lines in this diagram 
have to be put on-shell, which simplifies the calculation. 
Substituting free propagators
for the effective heavy-quark, light-quark
and gluon fields, we use Cutkosky rules and obtain
a dispersion relation for the Fig.~2 diagram contribution to the correlation
function
\begin{multline}
\Pi^{(\Gamma),pert}_\lambda(\rho,t,u)\:=\:\tilde{c}\int_0^\infty
\dfrac{ds}{s\,-\,\rho}\int\dfrac{d^4ld^4k}{(2\pi)^8}\,e^{-it(l+uk)\cdot
  n}(k_\lambda k_\alpha n_\beta\,+g_{\lambda\alpha} k_\beta\,n\cdot k)\\
\delta(s-(l\cdot v+k\cdot v))\delta(l^2)\delta(k^2)\Theta(l_0)\Theta(k_0)\,
\mbox{Tr}\left[\Gamma\dfrac{1+\fmslash{v}}{2}
\sigma^{\alpha\beta}\gamma_5\fmslash{\,l}\right]\,,
\label{eq-diag}
\end{multline}
where $l$ and $k$ are  
the four-momenta of the light-quark and gluon  
lines, respectively, and $\tilde{c}$ is the constant
factor (containing also $\alpha_s$). 
The integration is conveniently carried out 
if one expands these
momenta in light-cone components using the basis of the light-like
vectors $n$ and $\bar{n}$ introduced above: 
\begin{eqnarray}
k_\mu&=&\dfrac{1}{2}[(k\cdot\bar{n})\,n_\mu\,+\,(k\cdot n)\,
\bar{n}_\mu]\,+\,k_{\perp\mu}\,, 
\nonumber\\
l_\mu&=&\dfrac{1}{2}[(l\cdot\bar{n})\,n_\mu\,+\,(l\cdot n)\,
\bar{n}_\mu]\,+\,l_{\perp\mu}\,.
\label{eq-LCcomp}
\end{eqnarray}
The delta-functions in Eq.~(\ref{eq-diag}) are integrated out,
taking into account the kinematical bounds represented by $\Theta$-functions. At the end  two integrations are left, with the variables 
$\omega=(l\cdot n)$ and $\xi=(k\cdot n)$, that is, the plus 
components of the quark and gluon loop momenta, respectively. 
Matching the result of this calculation to 
the hadronic dispersion relation with the pole-term 
(\ref{eq-HQETpole}) and 
employing quark-hadron duality for the excited and continuum states  with an effective threshold $\tilde{s}_0$, 
we perform the Borel transformation.
Comparing the dependence on 
the variables $\omega$ and $\xi$ on both sides, 
the following sum rules for three-particles DA's
are obtained, in the perturbative loop approximation:
\begin{multline}
\Psi_A(\omega,\xi)\:=\:\Psi_V(\omega,\xi)\:=\: r\,\xi^2
\int\limits_{(\xi+\omega)/2}^{\tilde{s}_0}\,ds\,\,e^{(-s+\bar{\Lambda})/\tau}\\
\times (2s\,-\,\omega\,-\,\xi)^2\Theta(2\tilde{s}_0\,-\,\omega\,-\,\xi) +\ldots
\label{eq-SRpsiAV}
\end{multline}

\begin{multline}
X_A(\omega,\xi)\:=\: r\,\xi(2\omega-\xi)
\int\limits_{(\xi+\omega)/2}^{\tilde{s}_0}\,ds\,e^{(-s+\bar{\Lambda})/\tau}\\
\times (2s\,-\,\omega\,-\,\xi)^2\Theta(2\tilde{s}_0\,-\,\omega\,-\,\xi) +\ldots
\label{eq-SRXA}
\end{multline}

\begin{multline}
Y_A(\omega,\xi)\:=\: r\,\xi
\int\limits_{(\xi+\omega)/2}^{\tilde{s}_0}\,ds\,e^{(-s+\bar{\Lambda})/\tau}\\
\times (2s\,-\,\omega\,-\,\xi)^2\left( -\frac{(2s\,-\,\omega\,-\,\xi)}{3}
+3\omega -\xi\right)\Theta(2\tilde{s}_0\,-\,\omega\,-\,\xi) +\ldots
\label{eq-SRYA}
\end{multline}
with an equal coefficient  $r$ emerging from the constant factors
$C_\lambda$ and $\tilde{c}$ in Eqs.~(\ref{eq-HQETpole}) and (\ref{eq-diag}),
respectively. 
Importantly, in the loop 
approximation, $\Psi_V (\omega,\xi)$ and $\Psi_A(\omega,\xi)$
are equal, while $X_A$ and $Y_A$ have different forms. 
If one takes the local limit 
$t\to 0$ of the correlation function (\ref{eq-corrHQET}), the  
resulting two-point sum rules for the normalization constants 
of the DA's yield, $\int_0^\infty d\omega d\xi \Psi_A(\omega,\xi)=
\int_0^\infty d\omega d\xi \Psi_V(\omega,\xi)$. 

In fact, 
$\Psi_A(\omega,\xi)$ and $\Psi_V(\omega,\xi)$ have independent normalization 
conditions 
\begin{eqnarray}
\int_0^\infty\,d\omega\,\int_0^\infty\,d\xi\,\Psi_A(\omega,\,\xi) & = & \dfrac{\lambda_E^2}{3}\nonumber\\
\int_0^\infty\,d\omega\,\int_0^\infty\,d\xi\,\Psi_V(\omega,\,\xi) & = & \dfrac{\lambda_H^2}{3}
\label{eq-normal}
\end{eqnarray}
where the constants $\lambda_E$ and $\lambda_H$ are determined by
the matrix elements of different local operators \cite{GN}.
The functions $X$ and $Y$ are normalized to zero, but can also
have different normalization coefficients.   

The differences for all four DA's manifest themselves
if in the sum rules (\ref{eq-SRpsiAV})-(\ref{eq-SRYA})  
one takes into account the condensate contributions (indicated by ellipses), 
suppressed by the inverse powers of the Borel scale $\tau$. 
In fact, in the correlation functions with nonlocal currents 
the usual approximation of local quark and gluon condensates
is too crude and models of nonlocal condensates are usually employed 
\cite{GN,BIK}.  
We have investigated only the local limit of the correlation function
(\ref{eq-corrHQET}) and the resulting sum rules  
for $\lambda_E$ and $\lambda_H$.
The condensate contributions
have indeed  different sizes in these sum rules, 
but their influence on the  normalization constants is 
moderate, as compared with the loop contribution. 
The details of this analysis will be presented elsewhere.
Note that the sum rules for $\lambda_E$, $\lambda_H$
derived in \cite{GN} are based on 
a different, ``nondiagonal'' correlator with one three-particle 
and one two-particle current. These sum rules predict  
\be
\lambda_E^2= (0.11\pm 0.06) ~\mbox{GeV}^2,~~~
\lambda_H^2=(0.18\pm 0.07)~\mbox{GeV}^2,~~ 
\label{eq-lambEH}
\ee
not very far from each other, indicating 
that the approximation $\lambda_E =\lambda_H$
which follows from (\ref{eq-SRpsiAV}) can be adopted
within the uncertainty intervals in Eq.~(\ref{eq-lambEH}).

The most important prediction of   
the sum rules (\ref{eq-SRpsiAV})-(\ref{eq-SRYA}), is the
behavior at $\omega,\xi\to 0$  given  
by  the perturbative loop  contribution:
\begin{eqnarray}
\Psi_A(\omega,\xi) \sim \Psi_V(\omega,\xi) \sim \xi^2 \,,\nonumber \\ 
X_A (\omega,\xi) \sim \xi (2\omega-\xi),~~~ Y_A\sim \xi.
\label{eq-behav}
\end{eqnarray}
In our previous paper \cite{KMO}, we followed a different, more
qualitative way, making a comparison between the  
pion and $B$-meson three-particle DA's in the asymptotic regime
and  obtained 
\be
(\Psi_A\,-\,\Psi_V)\sim 
(\lambda_E^2-\lambda_H^2)\omega\xi^2\,,
\label{eq-correct}
\ee 
which turns out to be a small correction, neglected here. 
This correction does not contradict the behavior indicated in 
Eq.~(\ref{eq-behav}) but cannot be simply extracted from the  
sum rules (\ref{eq-SRpsiAV}) without going beyond the loop approximation.

We suggest two models for the three-particle DA's. 
The first one is obtained from the sum rules 
(\ref{eq-SRpsiAV})-(\ref{eq-SRYA})
in the local duality (LD) $\tau\to \infty$ limit: 
\begin{eqnarray}
\Psi_A^{LD}(\omega,\xi)& =&  \Psi_V^{LD}(\omega,\xi) =
\left(\dfrac{35\lambda_E^2}{4\tilde{s}_0^4}\right)\xi^2
\left(1-\dfrac{\omega+\xi}{2\tilde{s}_0}\right)^3
\Theta(2\tilde{s}_0-\omega-\xi),
\nonumber\\
X_A^{LD}(\omega,\xi)& =& 
\left(\dfrac{35\lambda_E^2}{4\tilde{s}_0^4}\right)\xi(2\omega-\xi)
\left(1-\dfrac{\omega+\xi}{2\tilde{s}_0}\right)^3
\Theta(2\tilde{s}_0-\omega-\xi),
\nonumber
\\
Y_A^{LD}(\omega,\xi)& =& 
-\left(\dfrac{35\lambda_E^2}{16\tilde{s}_0^4}\right)\xi
\left(1-\dfrac{\omega+\xi}{2\tilde{s}_0}\right)^3
(2\tilde{s}_0-13\omega+3\xi)
\Theta(2\tilde{s}_0-\omega-\xi).\nonumber\\
\label{3partLD}
\end{eqnarray}
The  uniform constant factor in the above expressions is fixed 
by the normalization conditions (\ref{eq-normal}), and we assume that
$\lambda_E=\lambda_H$. Note that $X_A(\omega,\xi)$ and $Y_A(\omega,\xi)$ 
in Eqs.~(\ref{eq-SRXA}), (\ref{eq-SRYA}) are 
normalized to zero, as they should be.  

It is natural to combine the above three-particle DA's with    
$\phi^B_{\pm}(\omega)$ obtained
in the same local-duality limit 
from the HQET sum rule for a correlator 
of the nonlocal and local quark-antiquark currents:
\begin{eqnarray}
\phi_{+}^{B,LD}(\omega)=\frac{3\omega}{2\tilde{s}_0^2}\left(1-\frac{\omega}{2\tilde{s}_0}\right)
\Theta(2\tilde{s}_0\,-\,\omega)\,,
\nonumber\\
\phi_{-}^{B,LD}(\omega)=\frac{3}{2\tilde{s}_0}\left(1-\frac{\omega}{2\tilde{s}_0}\right)^2
\Theta(2\tilde{s}_0\,-\,\omega)\,,
\label{phiLD}
\end{eqnarray}
where $\phi^{B,LD}_{+}$ has already been derived in 
\cite{BIK}.

For the second model of the three-particle DA's we combine 
the  small $\omega,\xi$ behavior (\ref{eq-behav}) 
with an exponential fall-off:
\begin{eqnarray}
\Psi_A(\omega,\,\xi)& =& \Psi_V(\omega,\,\xi) \,=\, 
\dfrac{\lambda_E^2 }{6\omega_0^4}\,\xi^2 e^{-(\omega\,+\,\xi)/\omega_0}\,,
\nonumber\\
X_A(\omega,\,\xi)& = & \dfrac{\lambda_E^2 }{6\omega_0^4}\,
\xi(2\omega-\xi)\,e^{-(\omega\,+\,\xi)/\omega_0}\,,\nonumber\\
Y_A(\omega,\,\xi)& =&  -\dfrac{\lambda_E^2 }{24\omega_0^4}\,
\xi(7\omega_0-13\omega+3\xi)e^{-(\omega\,+\,\xi)/\omega_0}\,.
\label{eq-3partexp}
\end{eqnarray}
The analogous ansatz for the two-particle DA's was suggested in \cite{GN}:
\begin{eqnarray}
\phi_+^B(\omega) & = & \dfrac{\omega}{\omega_0^2}\,e^{-\frac{\omega}{\omega_0}}\,,\nonumber\\
\phi_-^B(\omega) & = & \dfrac{1}{\omega_0}\,e^{-\frac{\omega}{\omega_0}}\,,
\label{eq-GN}
\end{eqnarray}
so that $\lambda_B=\omega_0$.

After the models are formulated, it is important 
to check if they obey the constraints derived in 
\cite{Kawamura} (see also \cite{BF}) from the QCD 
equations of motion (adapted to HQET) in the form of two equations for 
the two-particle DA's:
\begin{eqnarray}
\omega\dfrac{d\phi_-^B(\omega)}{d\omega}\,+\,\phi_+^B(\omega) & 
= & I(\omega)\,,\nonumber\\
\left(\omega\,-\,2\bar{\Lambda}\right)\phi_+^B(\omega)\,
+\,\omega\phi_-^B(\omega) & = & J(\omega)\,,
\label{eq-DArel}
\end{eqnarray}
where $I(\omega)$ and $J(\omega)$ are the 'source' terms due to 
the three-particle DA's:
\begin{eqnarray}
I(\omega) & = & 2\dfrac{d}{d\omega}\int_0^\omega\,
d\rho\int_{\omega-\rho}^\infty\,
\dfrac{d\xi}{\xi}\dfrac{\partial}{\partial\xi}
\left[\Psi_A(\rho,\,\xi)\,- \,\Psi_V(\rho,\,\xi)\right]\,,\nonumber\\
J(\omega) & = & -2\dfrac{d}{d\omega}
\int_0^\omega\,d\rho\int_{\omega-\rho}^\infty\,
\dfrac{d\xi}{\xi}\left[\Psi_A(\rho,\,\xi)\,
+\,X_A(\rho,\,\xi)\right]\nonumber\\
& & -\,4\int_0^\omega\,d\rho\int_{\omega-\rho}^\infty\,
\dfrac{d\xi}{\xi}\dfrac{\partial\Psi_V(\rho,\,\xi)}{\partial\xi}\,.
\label{eq-J}
\end{eqnarray}
We immediately notice that $I(\omega)=0$ in both models
(\ref{3partLD}) and (\ref{eq-3partexp}). 
In other words, the relation \cite{BF} 
\be
\phi_-^B(\omega)=\int\limits_\omega^\infty d\omega
\frac{\phi_+^B(\omega)}{\omega}
\label{eq-WWrel}
\ee
does not receive gluon corrections in the approximation
adopted here. Importantly, within this approximation also 
$J(0)=0$, hence the behavior $\phi_{+}^B(\omega)\sim \omega$ 
at $\omega\to 0$ is not modified, contrary to 
a general expectation \cite{Kawamura}.   

In addition, due to the relations between the matrix
elements of local operators the moments of DA's have to 
fulfil
\cite{GN} the following equations: \footnote{Note that after including renormalization effects
in DA's, which so far have only been studied for $\phi_+^B$ \cite{NL},
the positive moments of DA's logarithmically diverge, 
and have to be regularized in some way.} 
\begin{eqnarray}
\langle \omega \rangle_+ & =&  \dfrac{4}{3}\bar{\Lambda},~~
\langle \omega \rangle_-  =  \dfrac{2}{3}\bar{\Lambda}\,,\nonumber\\
\langle\omega^2\rangle_+ & =&  2\bar{\Lambda}^2\,+
\,\dfrac{2}{3}\lambda_E^2\,+\,\dfrac{1}{3}\lambda_H^2 \,,\nonumber \\ 
\langle\omega^2\rangle_- & = & \dfrac{2}{3}\bar{\Lambda}^2\,+
\,\dfrac{1}{3}\lambda_H^2\,,
\label{eq-relmoments}
\end{eqnarray}
 where $\langle \omega^n \rangle_{\pm}\equiv \int_0^\infty d\omega
\phi_{\pm}(\omega)\omega^n$. 

For the DA's of the local-duality model (\ref{phiLD}) and (\ref{3partLD}),   
only the first equation in (\ref{eq-DArel}) is valid. 
Hence, the 2-particle DA's acquire corrections determined by 
the three-particle DA's:
\be
\phi_{\pm}^{B,LD}(\omega) \to \phi_{\pm}^{B,LD}(\omega)+
\delta\phi_{\pm}^{B,LD}(\omega)\,.
\label{eq-ansatz}
\ee
Instead of substituting Eq.~(\ref{eq-ansatz})
into  Eqs.~(\ref{eq-DArel}) and solving the system of 
two equations for the functions 
$\delta\phi_{\pm}^{B,LD}(\omega)$, it is easier to 
use the ansatz obtained in \cite{Kawamura}:
\begin{eqnarray}
\phi_+^{B,LD}(\omega) +\delta\phi_+^{B,LD}(\omega)& = & \phi_+^{B,WW}(\omega)\,
+\,\dfrac{\omega}{2\bar{\Lambda}}\Phi(\omega)\,,\nonumber\\
\phi_-^{B,LD}(\omega) +\delta\phi_-^{B,LD}(\omega)& = & \phi_-^{B,WW}(\omega)\,
+\,\dfrac{2\bar{\Lambda}\,-\,\omega}{2\bar{\Lambda}}\Phi(\omega)\,
+\,\dfrac{J(\omega)}{\omega}\,,
\label{eq-ansPhi}
\end{eqnarray}
where $\phi_+^{B,WW}(\omega)$ are the solutions of 
Eqs.~(\ref{eq-DArel}) without the 'source' terms $I(\omega)$ and $J(\omega)$
(in Wandzura-Wilsczek (WW) approximation):
\begin{eqnarray}
\phi_+^{B,WW}(\omega) & = & \dfrac{\omega}{2\bar{\Lambda}^2}\,
\Theta(2\bar{\Lambda}\,-\,\omega)\,,\nonumber\\
\phi_-^{B,WW}(\omega) & = & \dfrac{2\bar{\Lambda}\,
-\,\omega}{2\bar{\Lambda}^2}\,\Theta(2\bar{\Lambda}\,-\,\omega)\,,
\label{eq-WW}
\end{eqnarray}
and  $\Phi(\omega)$ has a complicated expression via $I(\omega)$ and 
$J(\omega)$ which can be found in  \cite{Kawamura}. 
We have calculated  $J(\omega)$ for the local-duality model  (\ref{3partLD})
and the resulting function $\Phi(\omega)$.  
The results for $\phi_\pm^{B,LD}(\omega) +\delta\phi_\pm^{B,LD}(\omega)$
obtained from Eqs.~(\ref{eq-ansPhi}) are shown in Fig.~3. 
The corrected DA's differ significantly from the WW-approximation,
in particular, $(1/\lambda_B)^{WW}=1/\bar{\Lambda}$ is shifted to 
$ 1/\lambda_B=1/\bar{\Lambda}+7\lambda_E^2/(2\bar{\Lambda}
\tilde{s}_0^2)$,
while the positive moments satisfy Eqs.~(\ref{eq-relmoments})
(by construction of the ansatz  (\ref{eq-ansPhi})).
After including the gluon corrections, the functions $\phi_\pm^{B}(\omega)$
become smoother and are shifted towards lower $\omega$'s, as expected.. 

Turning to the exponential model and substituting 
Eqs.~(\ref{eq-3partexp}) for $\Psi_A$ and $X_A$ into
Eq.~(\ref{eq-J}), we obtain:  
\be 
J(\omega)=
\frac{2\lambda_E^2}{3\omega_0^4}\omega(\omega-2\omega_0)e^{-\frac{\omega}{\omega_0}}\,.
\label{eq-Jexp}
\ee
In this case, if the conditions \cite{GN}: 
\be
\omega_0=\frac23 \bar{\Lambda}, ~~\lambda_E^2=\lambda_H^2=\frac32\omega_0^2=
\frac23 \bar{\Lambda}^2\,.
\label{eq-paramrel}
\ee
are satisfied, both equations in (\ref{eq-DArel})
can be solved, and the solution for the two-particle DA's
has the exponential form (\ref{eq-GN}).
Under the same conditions, the relations (\ref{eq-relmoments})
between the moments are also fulfilled, as already 
noticed in \cite{GN}. 
Hence, the three-particle DA's 
described by the exponential model (\ref{eq-3partexp}) 
do not induce additional corrections to 
the ansatz (\ref{eq-GN}). We conclude that 
the combination of
Eqs.~(\ref{eq-GN}) and (\ref{eq-3partexp}), together with the conditions (\ref{eq-paramrel}) form a selfconsistent model 
of two- and three-particle $B$ meson DA's.

\begin{figure}[ht]
\includegraphics[width=6cm]{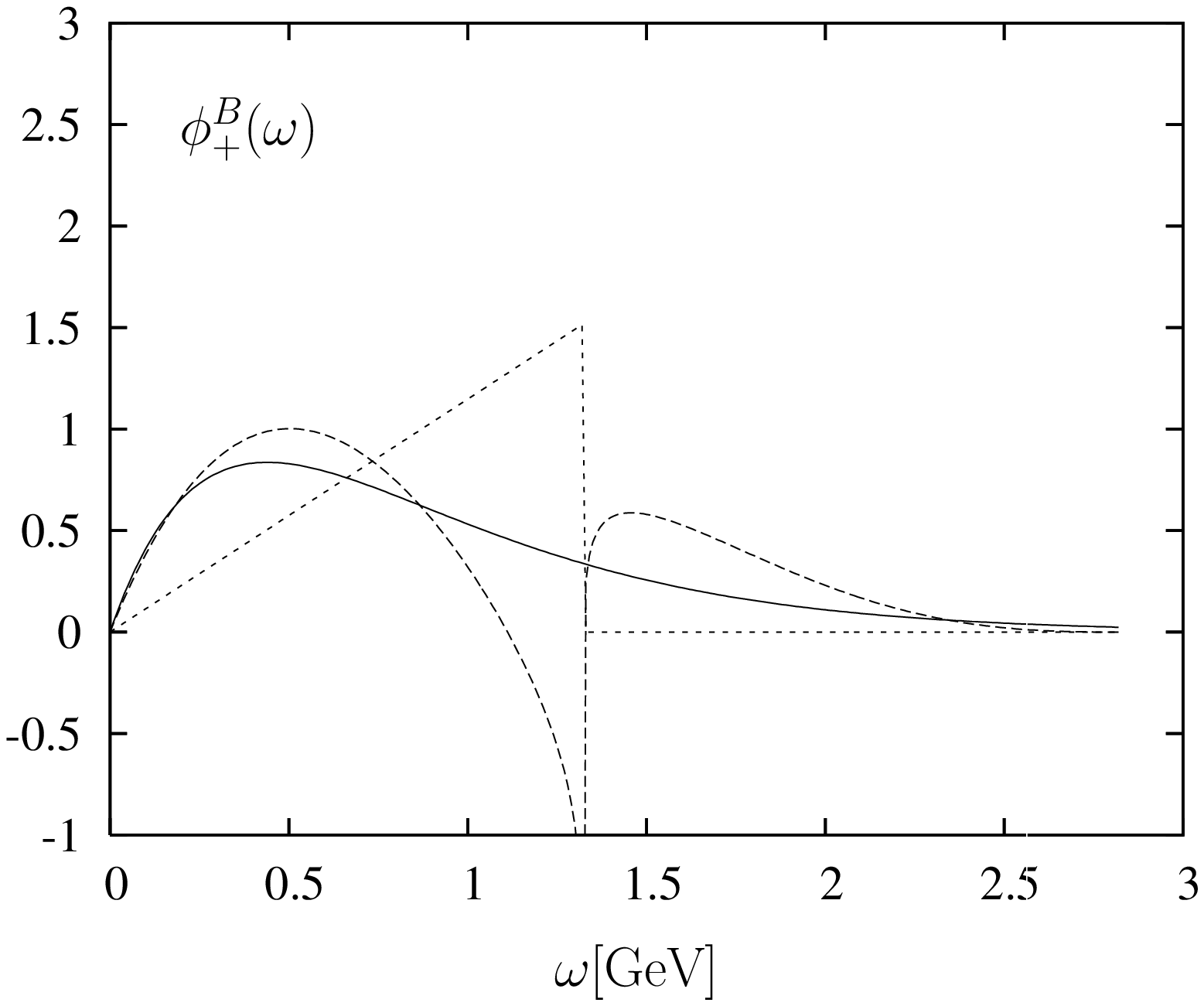}
\includegraphics[width=6cm]{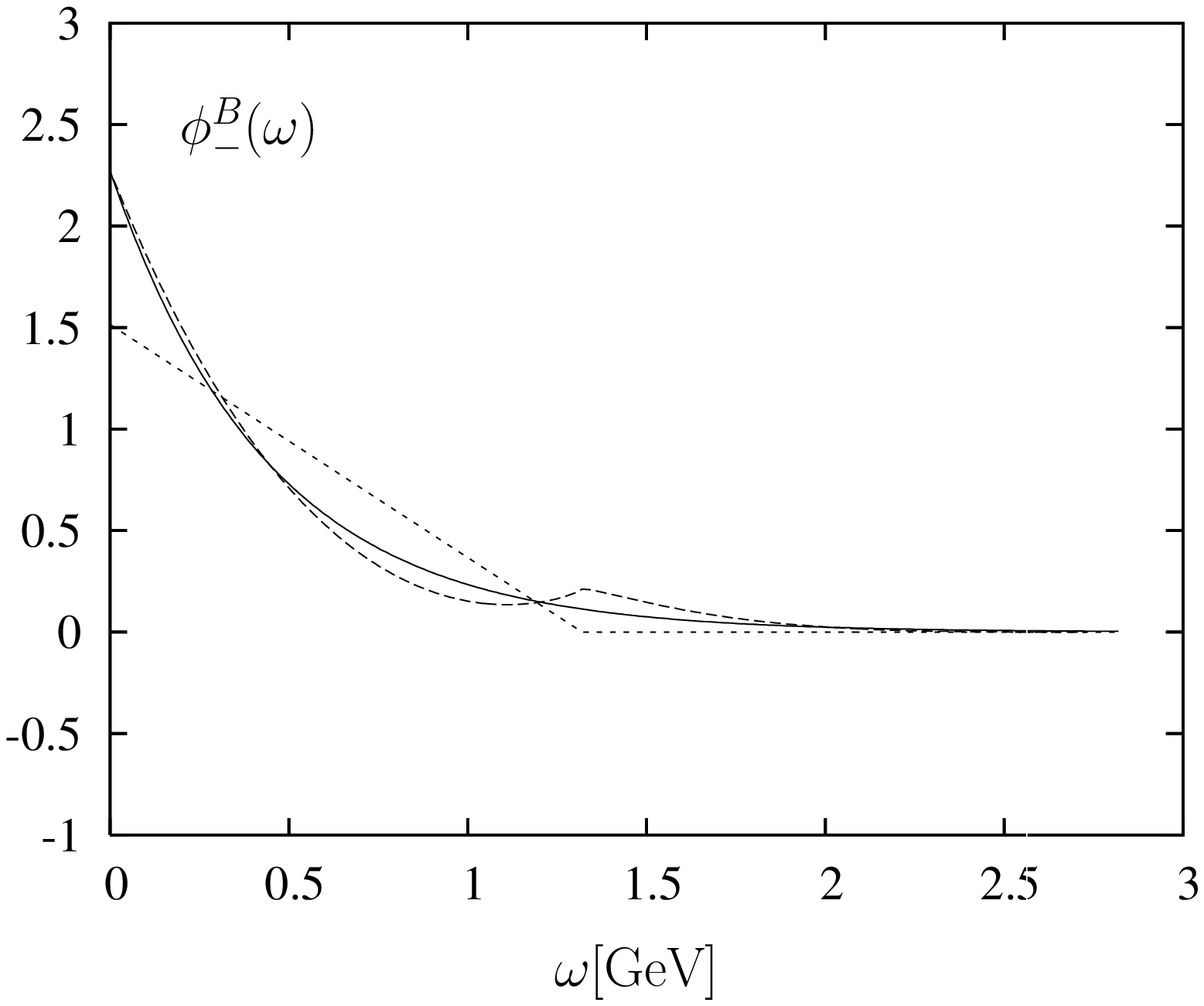}
\caption{ \it $B$-meson two-particle DA's $\phi_+^B(\omega)$ (left) 
and $\phi_-^B(\omega)$  (right). The solid (dashed) lines 
represent the  exponential (local-duality) model; 
short-dashed lines  are the WW-parts (\ref{eq-WW}).
In both models $\lambda_B\,=\,\frac{2}{3}\bar{\Lambda}$ 
and  $\lambda_E^2 =\lambda_H^2 = \frac{2}{3}\bar{\Lambda}^2$.}
\label{fig-DAmodels}
\end{figure}

Comparing the exponential model 
with the local-duality one introduced above, 
we find that numerically, 
in the region of integration in LCSR , 
$\omega<s_0^{\pi(K,\rho,K^*)}/m_B$, 
the two models for $\phi_{\pm}^B$ 
are almost indistinguishable 
(if $\lambda_E=\lambda_H$ and $\lambda_B$ are the same), 
as can be seen from Fig.~\ref{fig-DAmodels}. 
For that reason, in the numerical analysis of LCSR,  
we only consider the exponential model. 

In the region of small $\omega$ 
the exponential ansatz (\ref{eq-GN}) for $\phi_+^B(\omega)$ 
is numerically close to the 
more elaborated model suggested in \cite{BIK}: 
\be
\phi_+^B(\omega, \mu=1 \mbox{GeV})
=\frac{4\omega}{\pi \lambda_B(1+\omega^2)}
\left[\frac{1}{1+\omega^2}- \frac{2(\sigma_B-1)}{\pi^2}
\ln(\omega)\right]\,,
\label{eq-BIKmodel}
\ee
provided $\omega_0=\lambda_B$. In the above, 
$\omega$ is in GeV units and 
the parameters $\lambda_B$ and $\sigma_B$ are determined from 
HQET sum rules including the QCD radiative and nonperturbative 
corrections. 
Also the model for $\phi_+^B(\omega)$ suggested in 
\cite{LeeNeub} 
at small $\omega$  contains  the same exponential component 
as in  Eq.~(\ref{eq-GN}).

\section{ Heavy-mass limit of LCSR}

In this section we discuss the power counting in the 
$B$-meson LCSR, in particular, the dependence 
on the heavy-mass scale $m_B\sim m_b$ at $q^2=0$ 
(at large energies of the final $P$, $V$ mesons).

Let us remind that 
the concept of $B$-meson DA's in its present form 
is only valid in the framework of HQET. 
In deriving the sum rules, we actually started from 
the formal $1/m_b$ expansion
(\ref{eq-HQET}) of the correlation function (\ref{eq-corr}) 
and further used the HQET correlation function 
(\ref{eq-HQETcorr}) expanding it in $B$-meson DA's. Hence, beyond the 
adopted approximation, there remain some 
unaccounted $1/m_b$ corrections which 
contribute to the ``systematical'' uncertainty  
of our method. These corrections  can be studied 
by expanding both the heavy-light current and the $B$-meson state in 
Eq.~(\ref{eq-corr}) beyond the leading-order in HQET.

As explained in Sect.~2,3, the relevant scale 
in the light-cone OPE for the HQET correlation function 
is the virtuality $P^2$ 
in the light-meson channel, or the corresponding 
Borel parameter $M^2$. This scale is chosen to be 
large with respect to $\Lambda_{QCD}$ but is 
independent of $m_b$.
In LCSR for the pion form factors \cite{BKM,pionff,piongamma}
the higher-twist components of pion DA's, including the
3-particle (quark-antiquark-gluon) DA's, yield contributions that 
are normally suppressed by the inverse Borel scale.    
The absence of a well-defined twist  in $B$ meson DA's 
makes the situation for the LCSR obtained here quite different. The 
contributions of the three-particle DA's do not reveal 
a general $1/M^2$ suppression.
Note however, that only the leading-order terms 
of the $x^2$-expansion for  both quark-antiquark and 
quark-antiquark-gluon
matrix elements (\ref{eq-BDAdef}) and (\ref{eq-B3DAdef}) 
are taken into account. 
It is natural to expect that if one continues the light-cone expansion 
further, the $1/M^2$ hierarchy will emerge in full scale.

The main source of $1/m_b$ suppression in $B$-meson LCSR 
is related to the second large scale $m_b\omega$ 
present in the denominators of the correlation function
(see Eq.~(\ref{eq-inv})). In the sum rules 
(\ref{eq-fplBpi0})-(\ref{eq-T1Brho0})
this scale manifests itself in the dependence 
of DA's on $(s/m_B)$, bounded by the duality interval 
$s/m_B<s_0/m_B$.
The power of $1/m_b$ suppression is entirely determined 
by  the $\omega \to 0$ ($\omega, \xi\to 0 $) 
``infrared'' behavior of the $B$-meson two- (three-) particle DA's,
contributing to the sum rule.  
This resembles the $1/m_b$ expansion of the light-meson
LCSR where the end-point behavior of the pion or $\rho$-meson 
DA's provides additional  $1/m_b$ suppression.

Expanding  at $m_b\to \infty$ the LCSR 
(\ref{eq-fplBpi0})-(\ref{eq-T1Brho0}), and adding the 
three-particle contributions given in the Appendix, 
one easily recovers the well-known relations \cite{Charles} 
valid in the limit of the large light-meson energy ($E_{P,V}\sim m_b/2$):
\ba
&&f^+_{BP}(0)=\zeta, ~~~f^T_{BP}(0)=\left(1+\frac{m_P}{m_B}\right)\zeta\,, 
\nonumber \\
&&V(0)=\left( 1+\frac{m_V}{m_B}\right)\zeta_\perp ,~~~
A_1(0)=\frac{m_B}{m_B+m_V}\zeta_\perp\,,\nonumber\\
&&A_2(0)=\left(1+\frac{m_V}{m_B}\right)
\left(\zeta_\perp-\frac{2m_V}{m_B}\zeta_\parallel\right)\,, ~~~
T_1(0)=\zeta_\perp ,
\label{eq-LEET}
\ea
For the first two universal form factors the following expressions 
in terms of $B$-meson DA's are obtained:  
\ba 
\zeta&=&\frac{\hat{f}_B}{f_P m_B^{3/2}}e^{m_P^2/M^2}\int\limits_0^{s_0^P}ds e^{-s/M^2}\phi^B_-(0)\,,\label{eq-zeta}\\
\zeta_\perp&=&\frac{\hat{f}_B}{2f_V m_V m_B^{3/2}}e^{m_V^2/M^2}\int\limits_0^{s_0^V}ds\, e^{-s/M^2}\left\{s
\frac{d \phi^B_+(\omega )}{d\omega}\Big|_{\omega=0}
\right.\nonumber\\
&&\left.-\int_0^\infty\dfrac{d\xi}{\xi}\Psi_V(0,\xi)\right\}\,,
\label{eq-zetaperp}
\ea
where the $B$-meson decay constant is rescaled 
in a standard way:  $f_B=\hat{f}_B/\sqrt{m_b}$. 
In the above, we neglected the light-quark 
masses $m_{1,2}$ but left $m_P\neq 0 $ for generality.    
In deriving Eq.~(\ref{eq-zetaperp}) we have also  
taken into account that the integral 
contributing to r.h.s.,
$$\int_0^\infty\dfrac{d\xi}{\xi}\left(\Psi_A(0,\xi)+X_A(0,\xi)\right)=-\frac12J(0)=0$$
in our model.  
The third universal form factor $\zeta_\parallel$ enters  
Eq.~(\ref{eq-LEET}) for $A_2$ with an $O(1/m_B)$ factor, 
hence, it cannot be cleanly separated 
from the other $1/m_b$ corrections to the LCSR 
(\ref{eq-A2Brho0}) for $A_2$.  
One has to obtain a separate sum rule for $A_0$, 
but we do not dwell on that here.

The $1/m_b^{3/2}$ limit
for all form factors,  evident from Eqs.~(\ref{eq-zeta}),
(\ref{eq-zetaperp}), is consistent with the heavy-mass limit
obtained from the light-meson LCSR. The only  
exception is the heavy-mass limit 
$f_{B\pi}^{+}+ f_{B\pi}^{-}\sim 
O(1/m_b^{5/2})$, obtained from Eq.~(\ref{eq-fpmBpi0}),
and different from the $1/m_b^{3/2}$-behavior 
predicted from LCSR with the pion DA's \cite{KRW}.

Our main observation is that the universal 
$B\to P$ form factor $\zeta$ does not receive contributions from 
the three-particle $B$ meson DA's, 
while for the $B\to V$ form factors 
the three-particle Fock components in the $B$-meson 
contribute at the leading power $O(1/m_b^{3/2})$ 
with a universal term. This result agrees with the  
expectations of SCET  discussed in \cite{Lange}.
Also in the factorization formula for the form factor 
$\zeta$ derived in \cite{MSt} the quark-antiquark-gluon DA's  
contribute at the leading order.

\section{Numerical results}

To perform the numerical analysis of the new LCSR, 
we use the exponential model 
(\ref{eq-3partexp}), (\ref{eq-GN}) of $B$-meson DA's
and adopt 
the interval \cite{BIK}:
\be
 \lambda_B( 1~ \mbox{GeV})=460\pm 110 ~\mbox{MeV}
\label{eq-interv1}
\ee
for the inverse moment of $\phi_+^B$.  
The parameters $\lambda_E^2=\lambda_H^2$ 
are determined from Eq.~(\ref{eq-paramrel}), somewhat larger 
than in Eq.~(\ref{eq-lambEH}). 
In addition, having in mind 
the uncertainty of the model, we allow the 
parameters $\lambda_E=\lambda_H$ to vary within 
$\pm 50\%$ at fixed $\lambda_B$, so that the constraints 
following from equations of motion remain valid. 
The $B$-meson decay constant 
$f_B=180\pm 30$ MeV obtained from the two-point sum rule 
in $O(\alpha_s)$  is used, similar to \cite{BPLCSR}. 
This is consistent with the $O(\alpha_s)$ accuracy 
of $\lambda_B$. This matching of precisions is however 
not yet complete, in the absence of the $O(\alpha_s)$ 
corrections to LCSR.    

The interval of Borel parameter adopted here, $M^2=1.0\pm 0.5 $ GeV$^2$, 
is optimal for the two-point sum rules in the light-meson 
channels \cite{SVZ,CK}, as well as for LCSR for the 
pion form factors \cite{BKM, pionff,piongamma}. 
Hence, the normalization scale of 
$\lambda_B$ is consistent  with the average
virtuality in the correlation function.   
\begin{table}[h]
\begin{centering}
\begin{tabular}{|c|l|c|}
\hline
Meson & Decay constant \cite{PDG} & Threshold parameter \\
\hline
&&\\[-2mm] 
$\pi$ & $f_\pi= 130.7\pm 0.1$  MeV & 
$s_0^\pi= 0.7$ GeV$^2$ \cite{SVZ,CK}\\[-2mm]  
&&\\[-2mm] 
\hline
&&\\[-2mm] 
$K$ & $f_K= 159.8\pm 1.4 \pm 0.44$  MeV & $s_0^K=1.05$ GeV$^2$ \cite{KMM}\\ [-2mm]
&&\\[-2mm] 
\hline
&&\\[-2mm] 
$\rho$ & $f_\rho=209 \pm 2$  MeV & $s_0^\rho= 1.6$ GeV$^2$  \cite{SVZ,CK}
\\[-2mm] 
&&\\[-2mm] 
\hline
&&\\[-2mm] 
$K^*$ & $f_{K^*}= 217 \pm 5 $  MeV & $s_0^{K^*}= 1.7$ GeV$^2$  
\cite{BallZw3}\\ [-2mm]
&&\\
\hline
\end{tabular}\\
\caption{\it
Decay constants of light mesons 
and the threshold parameters extracted from
the corresponding 2-point QCD sum rules.}
\end{centering}
\label{tab-light1}
\end{table}
The input for various  light-meson channels 
is listed in Table~2. As already 
mentioned, the duality-threshold parameter in each channel 
is fixed by adjusting the two-point sum rule 
(taken with $O(\alpha_s)$ accuracy) to the 
experimentally measured decay constant. 
Note that the same values of $s_0^\pi$ and $s_0^\rho$ were used
in LCSR for the pion electromagnetic \cite{BKM,pionff} 
and $\rho\pi\gamma$, $\pi \gamma\gamma^*$ \cite{piongamma} 
form factors, respectively.
 
For the channels with strange mesons we adopt
$m_s( 1 \mbox{GeV}) =130\pm 10$  MeV which 
agrees, e.g., with the recent QCD sum rule estimates \cite{ms}.

\begin{figure}[ht]
\includegraphics[width=6cm]{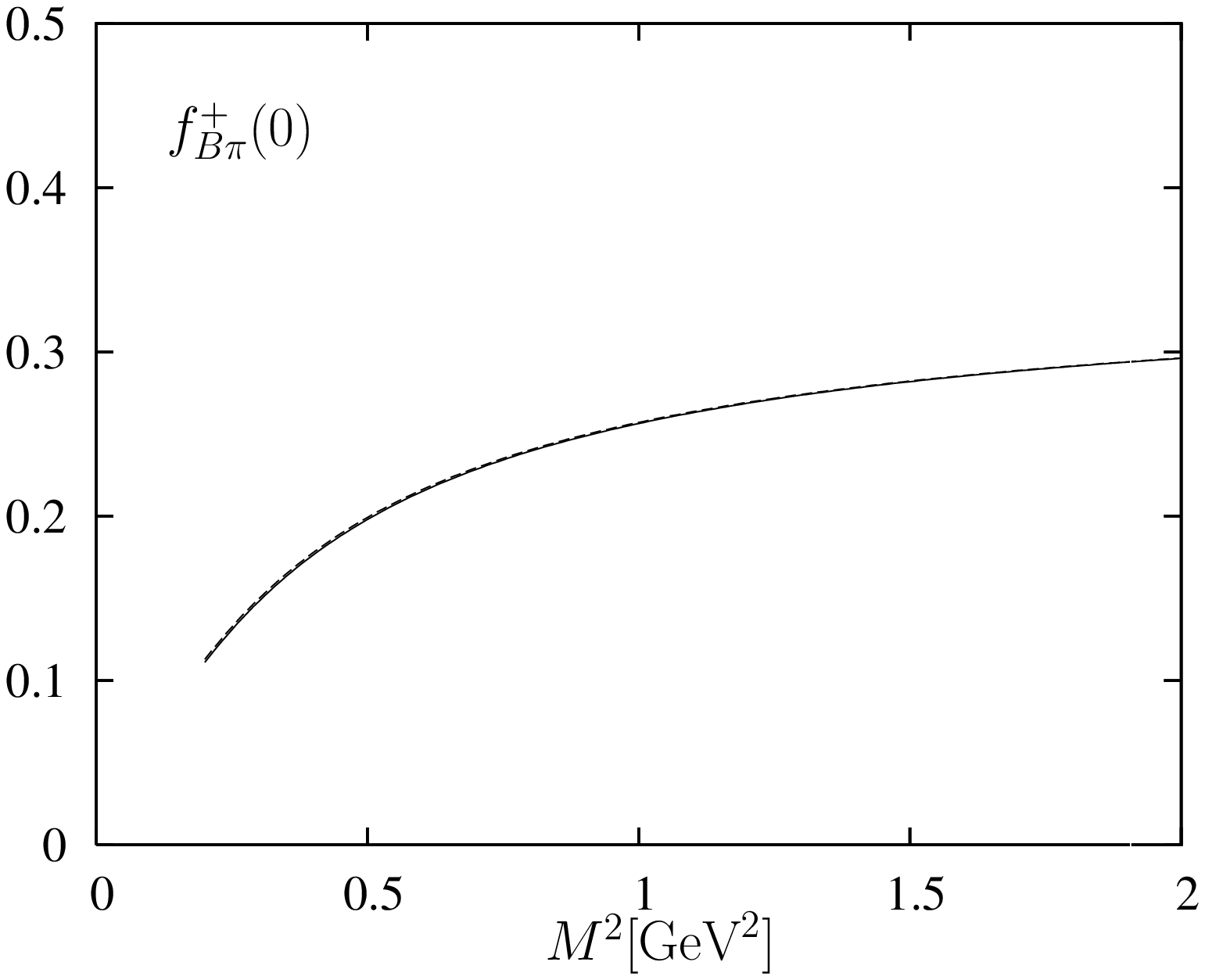}
\includegraphics[width=6cm]{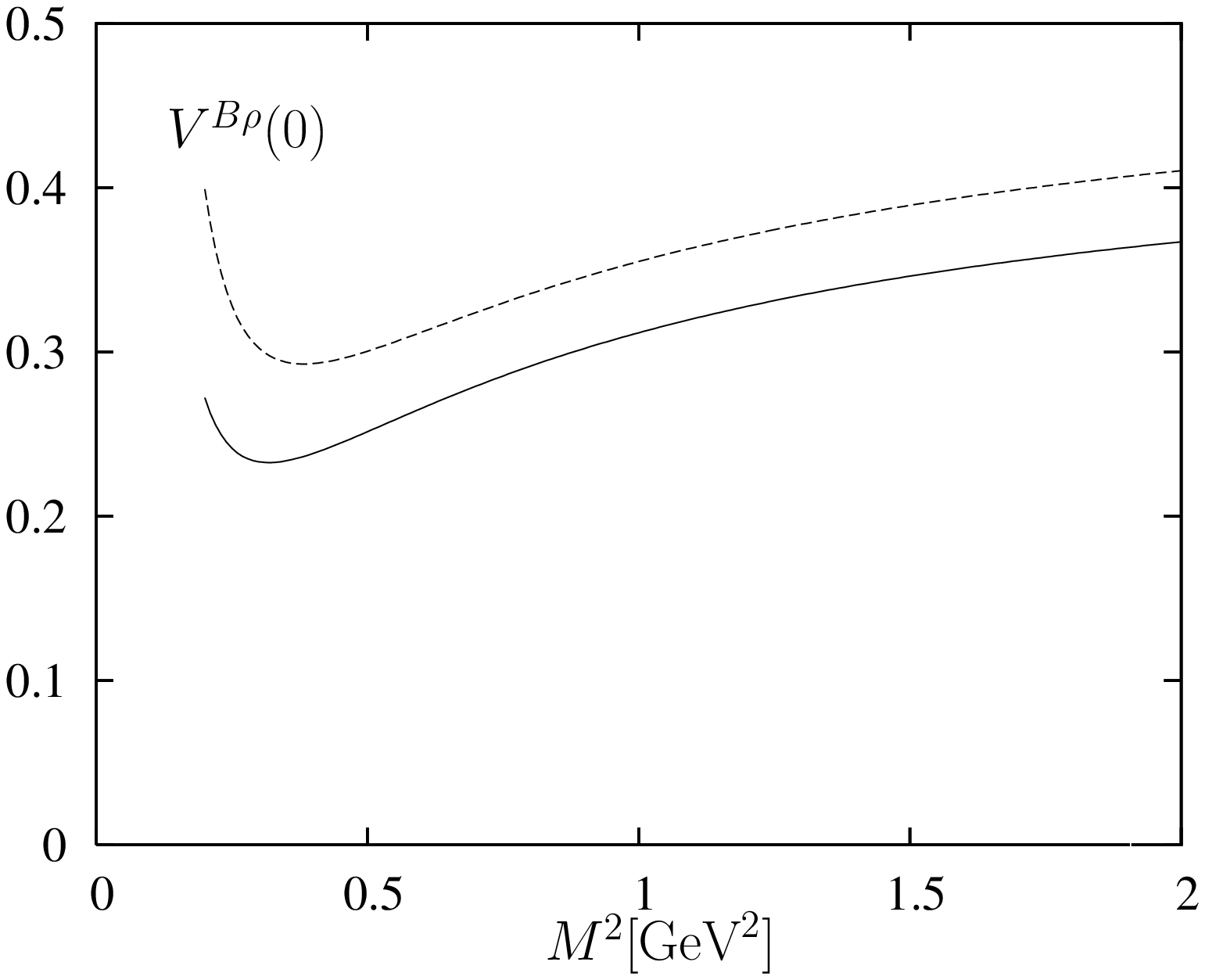}
\caption{\it Dependence of the form factors 
$f^+_{B\pi}(0)$ (left) and $V^{B\rho}(0)$ (right) 
on the Borel parameter (solid lines). The contributions of 
two-particle DA's are shown by dashed lines, almost 
indistinguishable from the total result for  $f^+_{B\pi}(0)$.}
\label{fig-Borel}
\end{figure}

\begin{figure}[h]
\begin{center}
\includegraphics[width=6cm]{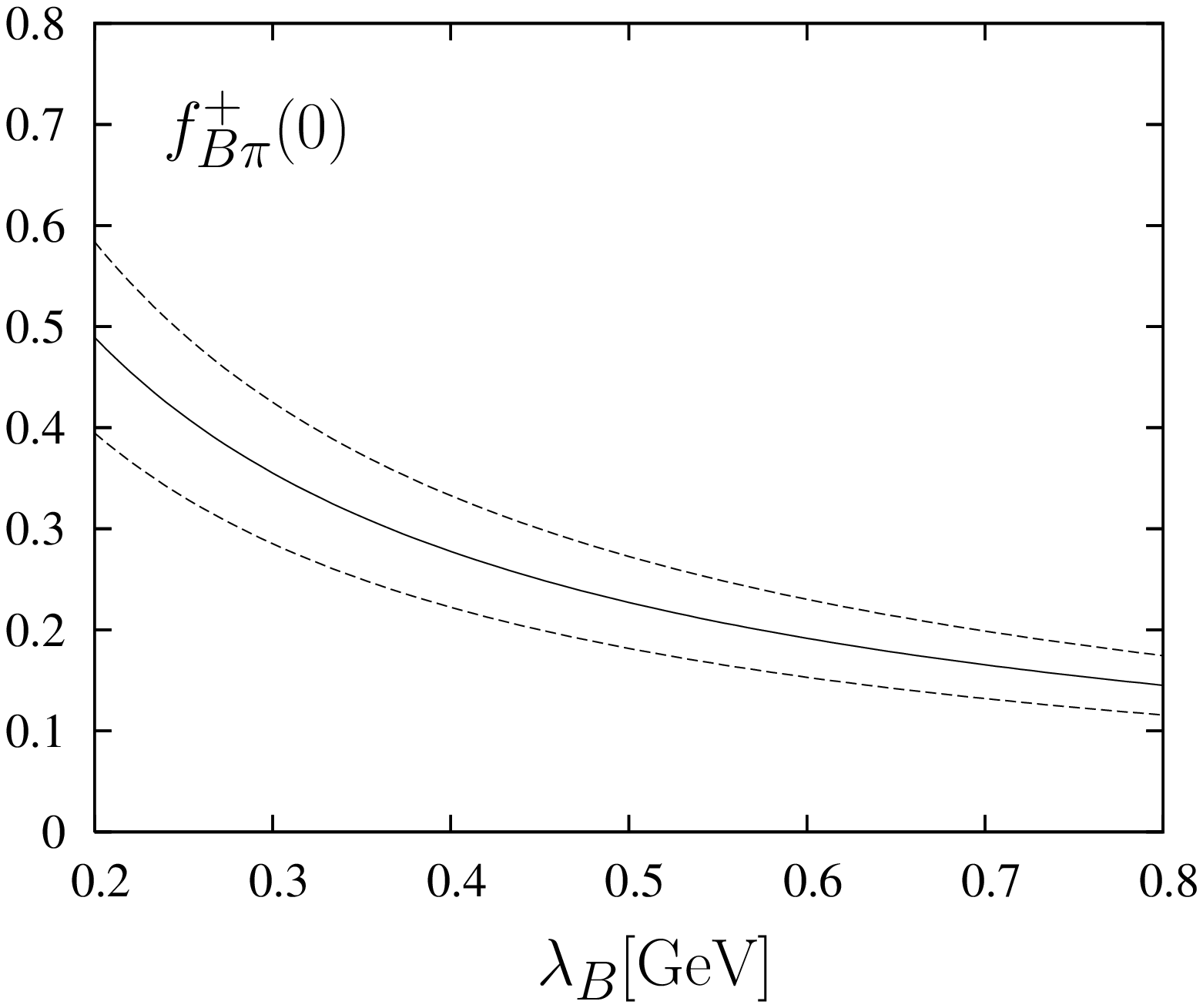}
\end{center}
\caption{\it Dependence of the form factor $f^+_{B\pi}(0)$
on the inverse moment $\lambda_B$. The solid line
corresponds to  the central values of all other input parameters. 
Their variation yields the uncertainty
interval shown with the dashed lines.}
\label{fig-fBpivslamB}
\end{figure}

To demonstrate the stability of the LCSR predictions 
with respect to the Borel parameter variation, 
as well as the role of three-particle corrections we plot
the numerical results for the two representative form factors
$f^+_{B\pi}(0)$ and  $V^{B\rho}(0)$ in Fig.~\ref{fig-Borel}.
The contribution of the three-particle DA's 
to the sum rule for $V^{B\rho}$ is substantially larger 
than the analogous contribution to the sum rule 
for $f^+_{B\pi}$; this observation  
is consistent with different $1/m_b$ behavior 
of the three-particle corrections, as 
discussed in the previous section. 
Furthermore, to illustrate 
the sensitivity of $B$-meson LCSR to the value of the 
inverse moment $\lambda_B$, we plot in 
Fig.~\ref{fig-fBpivslamB} our prediction for $f_{B\pi}(0)$ 
as a function of this input parameter.  

{\small
\begin{table}
\begin{centering}
\begin{tabular}{|c|c|l|}
\hline
&&\\[-1,5mm]
form factor & this work  & LCSR with light-meson DA's \\
\hline 
&&\\[-1,5mm]
$f^+_{B\pi}(0)$& 0.25$\pm$0.05 & 0.258$\pm$0.031\\ 
&&\\[-1,5mm]
\hline
&&\\[-1,5mm]
$f^+_{B K}(0)$& 0.31$\pm$0.04   & 0.301$\pm$0.041$\pm$0.008 \\
&&\\[-1,5mm]
\hline
&&\\[-1,5mm]
$f^T_{B\pi}(0)$& 0.21$\pm$0.04  & 0.253$\pm$0.028\\
&&\\[-1,5mm]
\hline
&&\\[-1,5mm]
$f^T_{B K}(0)$& 0.27$\pm$0.04 & 0.321$\pm$0.037$\pm$0.009\\
&&\\[-1,5mm]
\hline
&&\\[-1,5mm]
$V^{B \rho}(0)$& 0.32$\pm$0.10 & 0.323$\pm$0.029\\
&&\\[-1,5mm]
\hline
&&\\[-1,5mm]
$V^{B K^*}(0)$& 0.39$\pm$0.11& 0.411$\pm$0.033$\pm$0.031\\
&&\\[-1,5mm]
\hline
&&\\[-1,5mm]
$A_1^{B \rho}(0)$& 0.24$\pm$0.08& 0.242$\pm$0.024\\
&&\\[-1,5mm]
\hline
&&\\[-1,5mm]
$A_1^{B K^*}(0)$& 0.30$\pm$0.08 &  0.292$\pm$0.028$\pm$0.023\\
&&\\[-1,5mm]
\hline
&&\\[-1,5mm]
$A_2^{B \rho}(0)$& 0.21$\pm$0.09 & 0.221$\pm$0.023 \\
&&\\[-1,5mm]
\hline
&&\\[-1,5mm]
$A_2^{B K^*}(0)$& 0.26$\pm$0.08 & 0.259$\pm$0.027$\pm$0.022\\
&&\\[-1,5mm]
\hline
&&\\[-1,5mm]
$T_1^{B \rho}(0)$&0.28$\pm$0.09   & 0.267$\pm$0.021\\
&&\\[-1,5mm]
\hline
&&\\[-1,5mm]
$T_1^{B K^*}(0)$&0.33$\pm$0.10 & 0.333$\pm$0.028$\pm$0.024\\
&&\\
\hline
\end{tabular}
\caption{\it The $B\to \pi,K$ and $B\to \rho,K^*$ form factors 
calculated  in this work, compared with the predictions of 
the light-meson LCSR obtained in \cite{BallZw_BP}
and \cite{BallZw_BV}, respectively. For the latter,
the second uncertainty of
the $B\to K(K^*)$ form factors is due to the first 
Gegenbauer moment in the kaon ($K^*$) DA, 
where $a_1^K(\mbox{1~GeV})=0.05\pm 0.03$ 
($a_1^{K^*}(\mbox{1~GeV})=0.10\pm 0.07$) is taken. 
}
\end{centering}
\label{tab-res}
\end{table}
}
The form factors at zero momentum transfer 
calculated with the input specified above are collected in 
second column of Table~3. 
To estimate the theoretical uncertainties, one 
usually adds linearly or in quadrature the  uncertainties 
originating from separate  variations of 
the input parameters. 
The intervals presented in Table~3 are obtained 
with a different procedure. The central value for each 
form factor is fitted to the set of LCSR predictions 
obtained by simultaneosly scanning all input parameters 
($\lambda_B$, $\lambda_{E,H}^2$, $f_B$, $M^2$, $f_{P,V}$
and $m_s$) within the 
adopted ranges. The errors attributed to the fitted values 
are the usual 1$\sigma$ deviations.
The estimated uncertainties to a large extent 
originate from the interval of $\lambda_B$ ,
hence they are larger for $B\to V$ form factors
than for the $B\to P$ form factors,
because the former (latter) mainly depend on 
$\phi^{B'}_+(0)\sim \lambda_B^{-2}$
(on $\phi^B_-(0)\sim \lambda_B^{-1}$).
Simultaneosly, the ratios of the form factors have much smaller
uncertainties, in other words, the variations within the 
intervals presented 
in Table~3 are correlated. For example, the lower (upper) 
boundary
of the interval for $T_1^{B\rho (BK^*)}(0)$  
corresponds to the lower (upper) boundaries for $V^{B\rho(BK^*)}(0)$,
$A_1^{B\rho,(BK^*)}(0)$ and $A_2^{B\rho (BK^*)}(0)$. 

In Table~3 the predictions of the $B$-meson LCSR  are 
compared with the form factors obtained 
\cite{BallZw_BP,BallZw_BV}
from the conventional light-meson LCSR. One has to keep in mind 
that the latter sum rules are more precise,
because they include NLO corrections and are based
on the well-developed twist expansion. Hence,
the observed agreement between the predictions 
of two different methods is encouraging, possibly 
indicating that the unaccounted 
$O(\alpha_s)$ and power corrections 
to the new $B$-meson LCSR are  not large.
\begin{figure}[ht]
\includegraphics[width=6cm]{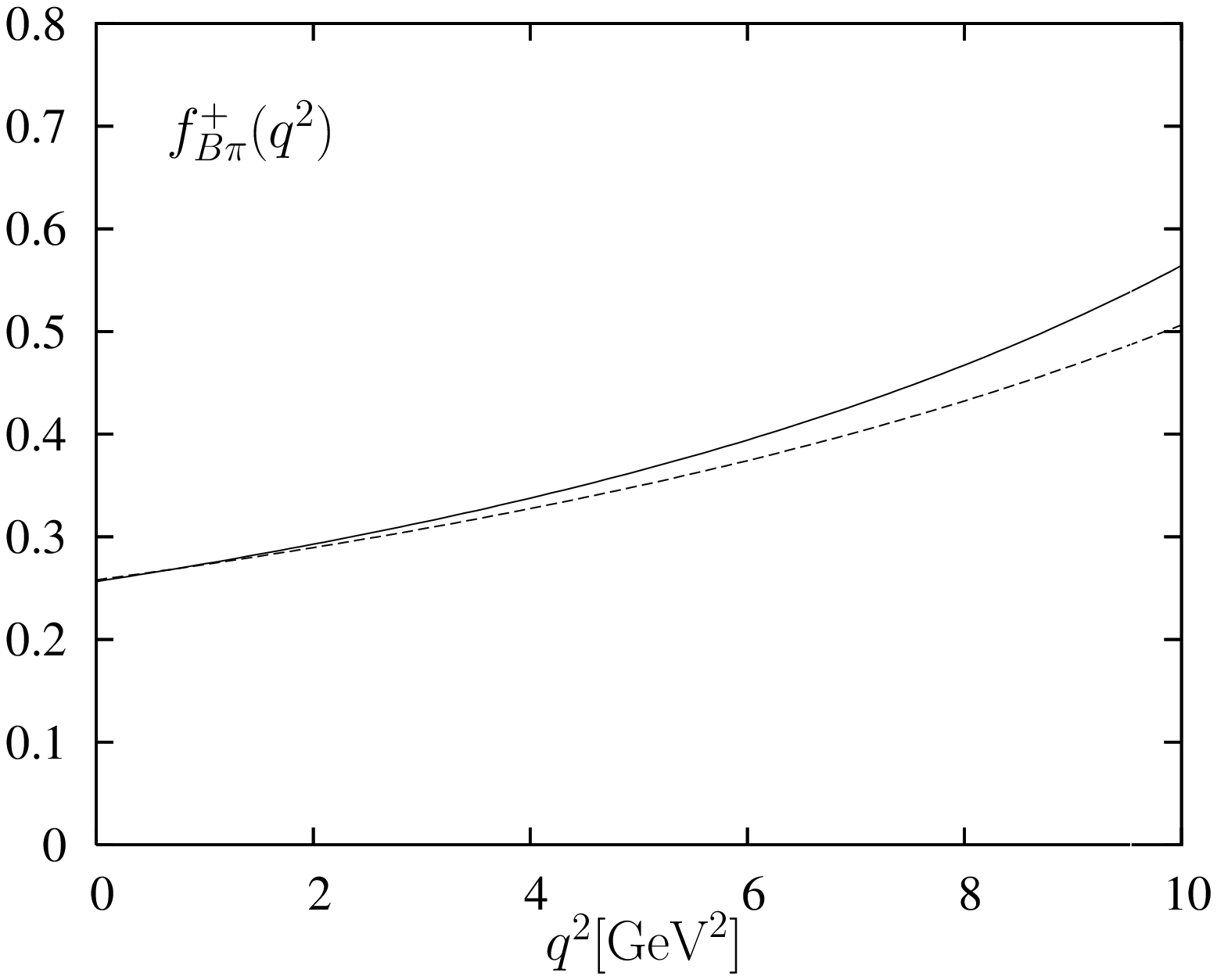}
\includegraphics[width=6cm]{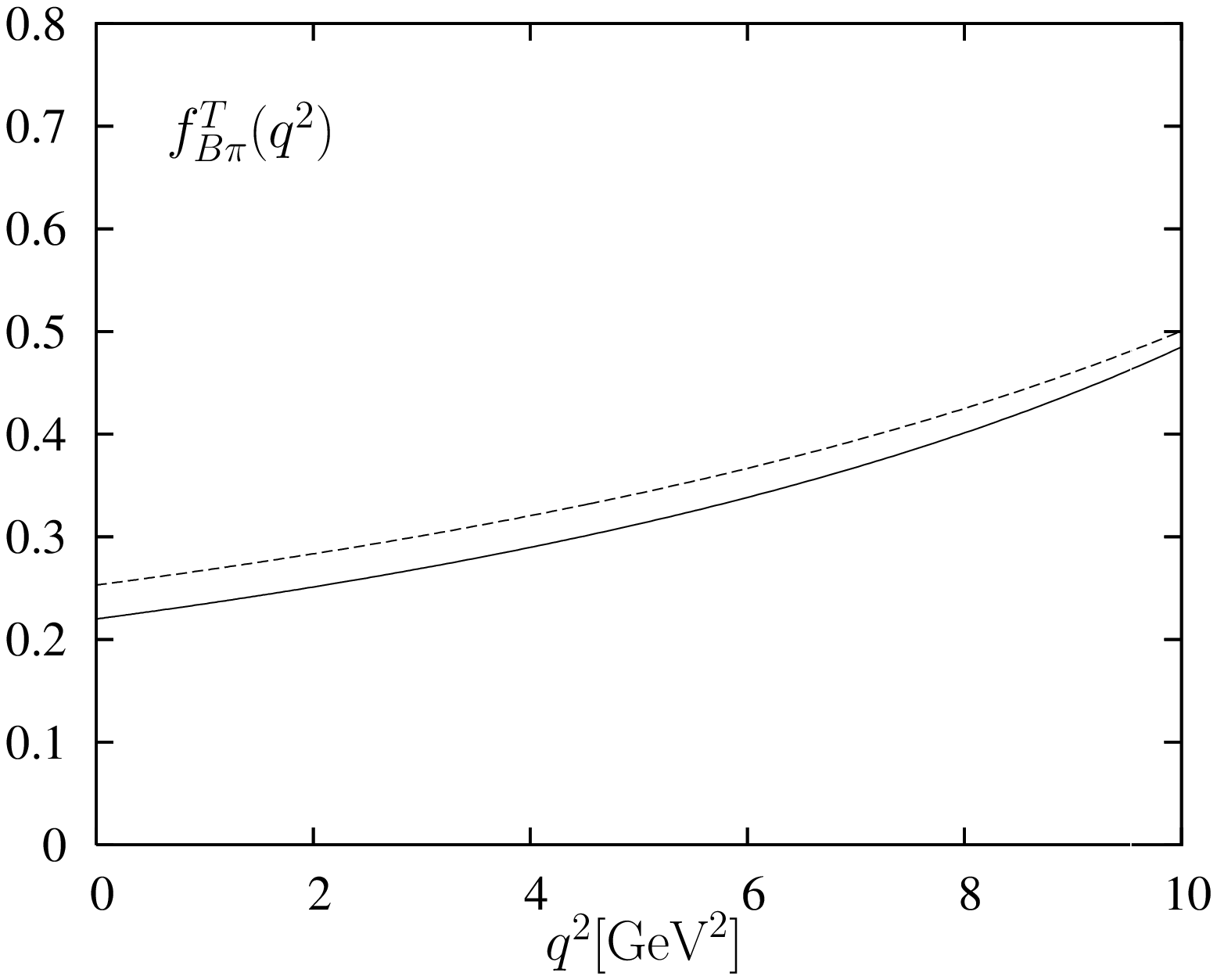}\\
\includegraphics[width=6cm]{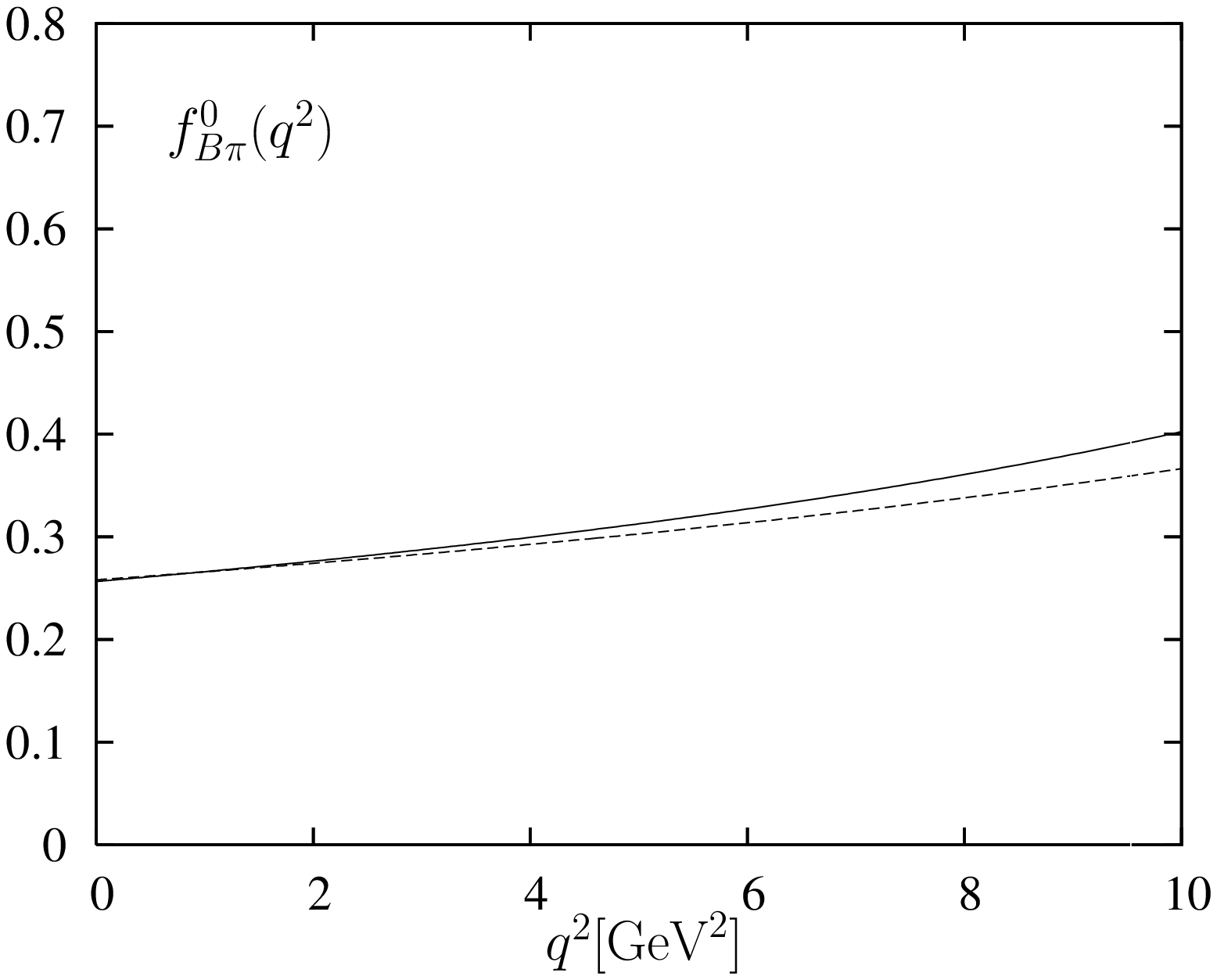}
\caption{\it Dependence of the $B\to \pi $ form factors on the momentum 
transfer squared (solid lines)
compared with the fits to the light-meson LCSR predictions 
from \cite{BallZw_BP} (dashed lines).
The 
theoretical uncertainties are not shown.}
\label{fig-q2depP}
\end{figure}

\begin{figure}[ht]
\includegraphics[width=6cm]{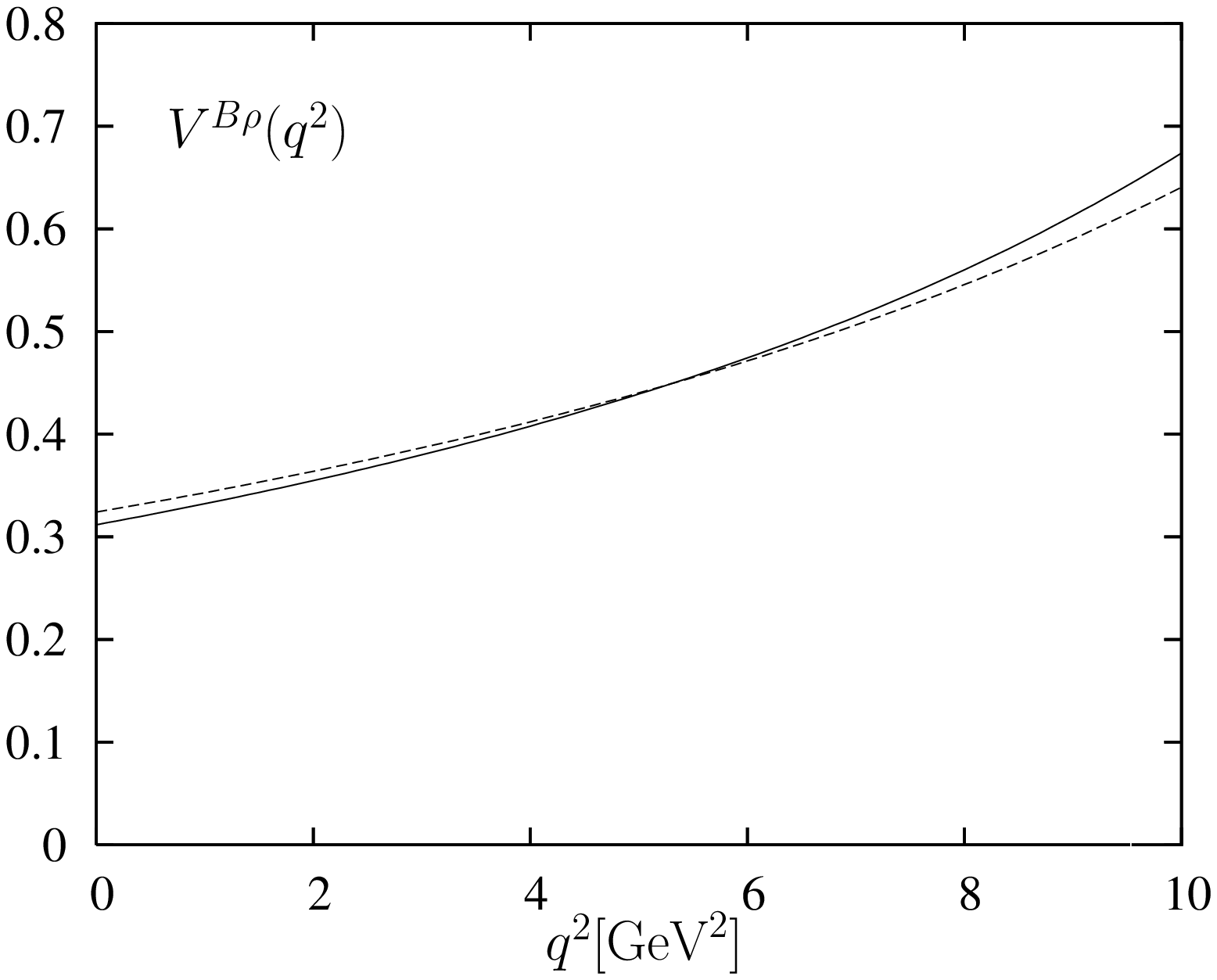}
\includegraphics[width=6cm]{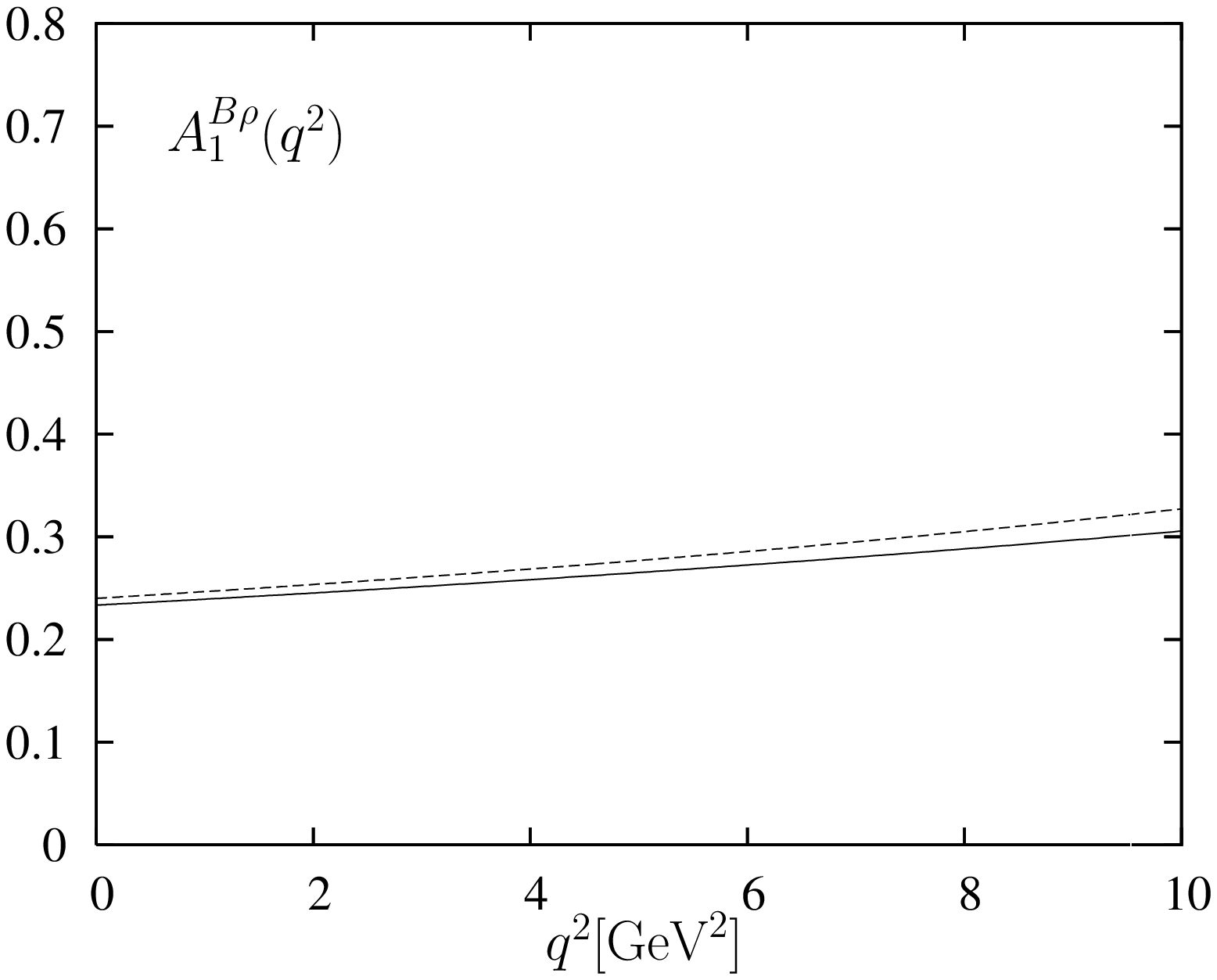}\\
\includegraphics[width=6cm]{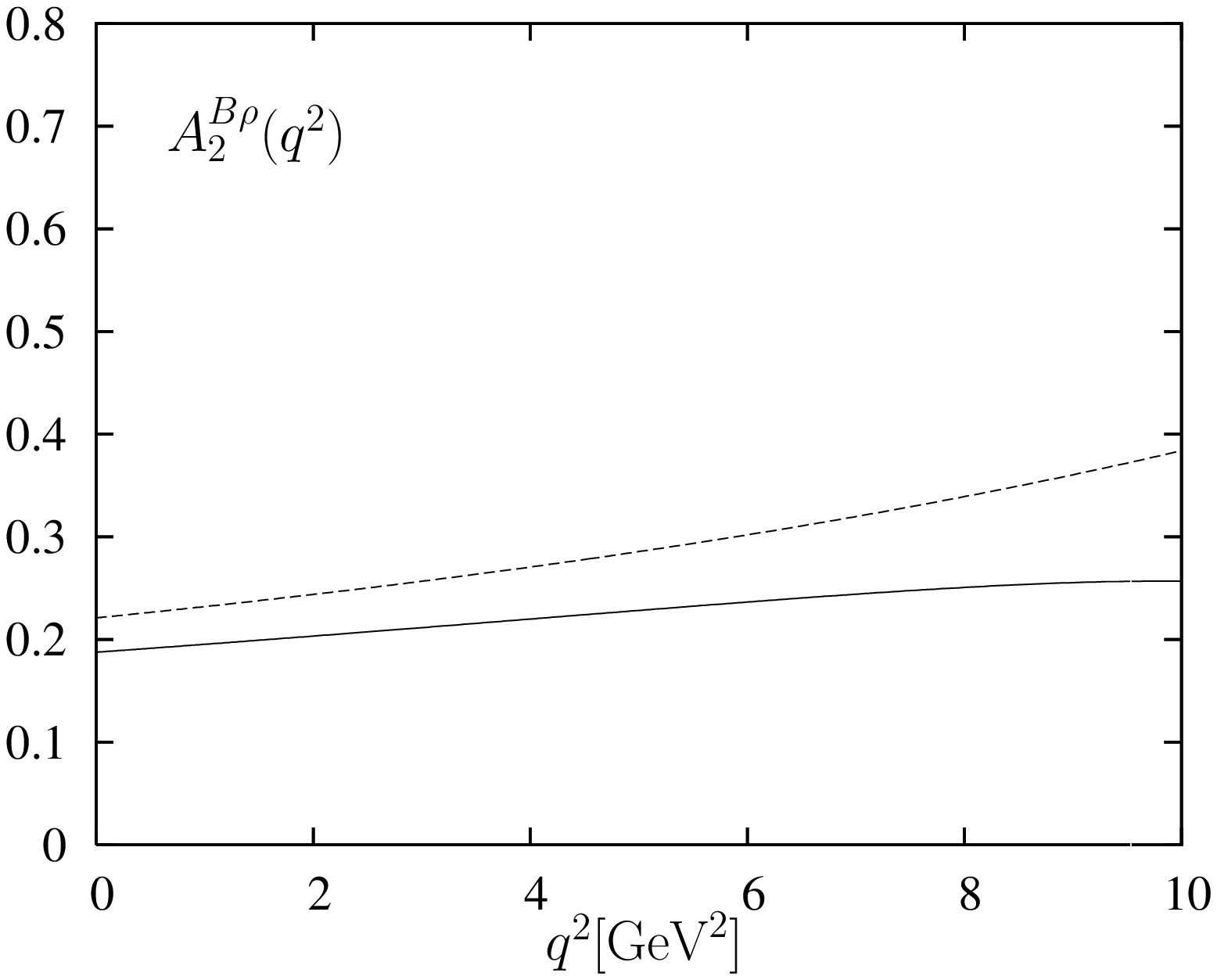}
\includegraphics[width=6cm]{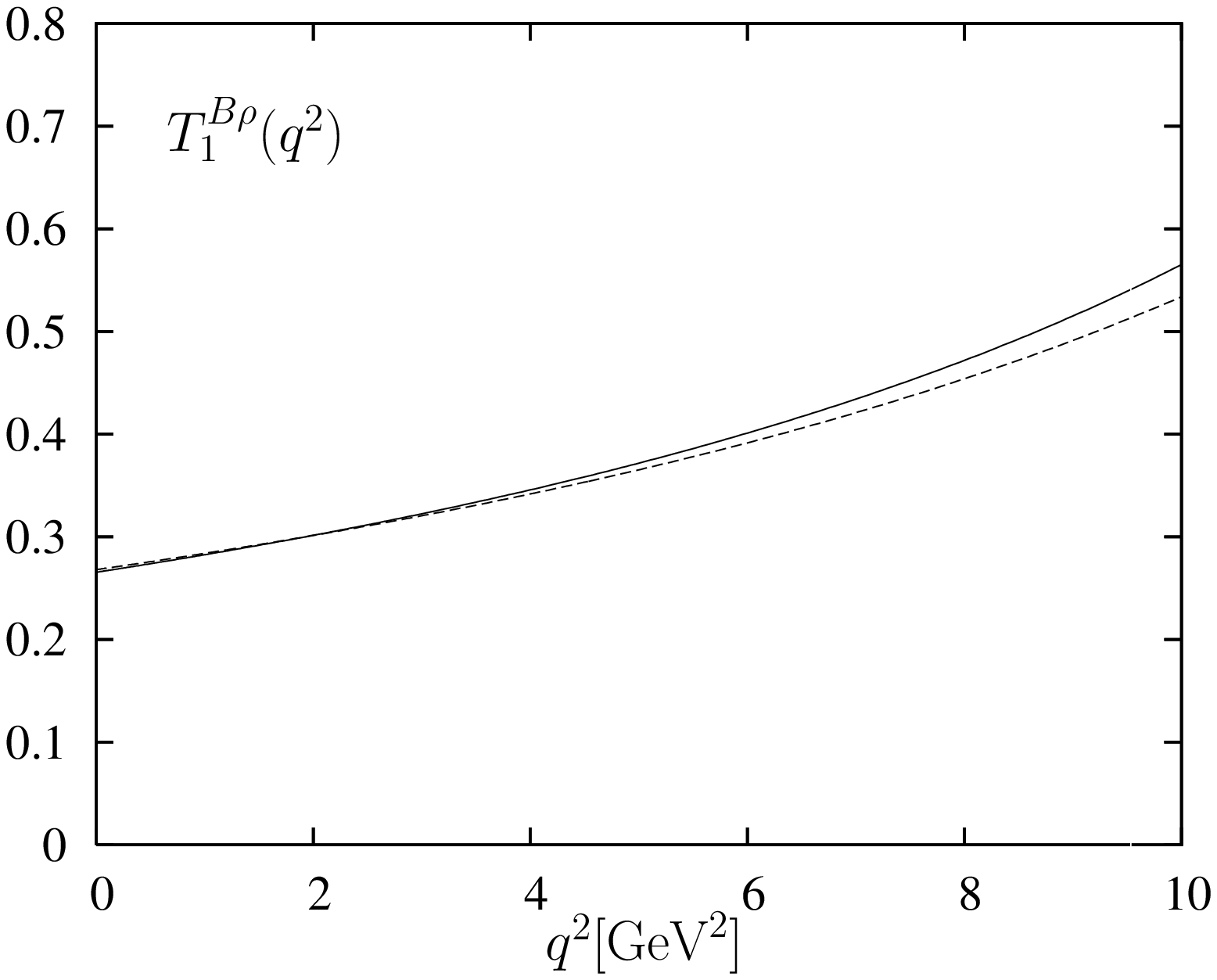}\\
\caption{\it The same as in Fig.~6, but for the $B\to \rho $ 
form factors; the light-meson LCSR results 
are from \cite{BallZw_BV}.}
\label{fig-q2depV}
\end{figure}

One more comment is in order.
As already mentioned, in the $B$-meson and light-meson LCSR,
quark-hadron duality is employed differently, 
in the light-meson and $B$-meson channels, 
respectively. 
Hence, the difference  between the predictions 
of two LCSR for one and the same form factor can be  
interpreted as a quantitative estimate 
of the ``systematic'' uncertainty caused by the 
duality ansatz. With the current accuracy of the $B$-meson 
LCSR we can only assess the upper limits 
for such uncertainties, by comparing the form factors 
in the second and third column of Table~3. 
To substantiate these estimates, one has to 
enhance the accuracy of the $B$-meson LCSR, e.g., by 
including the perturbative corrections and narrowing the 
ranges of the major inputs parameters, such as $\lambda_B$.

The $B\to \pi,\rho$  form factors   
calculated at $q^2\neq 0$  are plotted in 
Figs.~\ref{fig-q2depP},~\ref{fig-q2depV}. 
Note that to obtain the scalar form factor 
$f^0_{B\pi}(q^2)$ we have simply combined our predictions for 
$f^+_{B\pi}(q^2)$ and $f^+_{B\pi}(q^2)+f^-_{B\pi}(q^2)$.    
We evaluate the form factors at $0<q^2<10$ GeV$^2$.  
The adopted range corresponds to the expected validity 
interval (\ref{eq-interv}) where the upper boundary is taken at 
$P^2\sim M^2$. The light-meson LCSR are applicable
\cite{BPLCSR,BallZw_BP} at larger momentum transfers, 
up to 14-16 GeV$^2$. Note that the intervals presented 
in Table~3 
include theoretical uncertainties,
estimated as explained above, whereas the solid lines 
displayed in  Figs. ~\ref{fig-q2depP},~\ref{fig-q2depV}
have been evaluated at the central values of all parameters, 
hence, at $q^2=0$ these lines slightly deviate 
from the central points of the intervals. 
The estimated uncertainties of the form factors 
at $q^2\neq 0$,  not shown here, 
are at the same level and have the same
correlations as the uncertainties at $q^2 = 0$ discussed above. 

Comparison with  the results 
of the light-meson LCSR \cite{BallZw_BP,BallZw_BV}
reveals an agreement also at $q^2\neq 0$, as can be seen 
from Figs.~\ref{fig-q2depP},~\ref{fig-q2depV}, taking into
account also the uncertainties of both  methods. 
To quantify the difference between the 
predictions of the two different types of LCSR  at $q^2\neq 0$, 
we have fitted our results at $0<q^2<10 $ GeV$^2$ 
to the parametrization \cite{BecK}
used in \cite{BallZw_BP,BallZw_BV}, e.g., for the 
$f_{B\pi}^+(q^2)$ form factor we employ:
\be
f_{B\pi}^+(q^2)=\frac{r_1}{1-q^2/m_{B^*}^2}+\frac{r_2}{1-q^2/m_{fit}^2}\,.
\label{eq-fit}
\ee
Since the adopted range of $q^2$  is rather narrow 
and the theoretical uncertainties  are relatively large, it is 
difficult to fit all three parameters in 
Eq.(\ref{eq-fit}) without producing an unphysically low 
mass $m_{fit}$ of the second (effective) pole.  The situation 
is improved
if one adopts the value of $m_{fit}=6.38$ GeV  
from \cite{BallZw_BP}  and fits only the residues of the two poles
in  (\ref{eq-fit}). The result: $r_1=0.93$ and $r_2=-0.68$  
has to be compared with $r_1=0.744$ and $r_2=-0.486$ obtained in 
\cite{BallZw_BP}. 
The analogous fit of $B$-meson LCSR for the form factor 
$V^{B\rho}(q^2)$ yields (again at the fixed \cite{BallZw_BV}
mass $m_{fit}=6.19$ GeV) the residues:  
$r_1=1.10$ and $r_2=-0.80$, very close to  
$r_1=1.045$ and $r_2=-0.721$ obtained in \cite{BallZw_BV}. 
A more detailed study of parameterizations at $q^2\neq 0$, 
including all  $B\to P,V$ form factors, 
as well as the applications to heavy-light semileptonic 
and radiative decays, are subjects of a future study.

Furthermore, to illustrate 
the $SU(3)$-violation effects predicted 
from the $B$-meson LCSR, we have calculated the ratios:
\be
\frac{f^+_{B K}(0)}{f^+_{B \pi}(0)}=1.27\pm 0.07 \,,
\label{fplratio}
\ee 
\be
\frac{T_1^{B K^*}(0)}{T_1^{B \rho}(0)}=1.22\pm 0.13\,. 
\label{Tratio}
\ee 
Importantly, these ratios are much less dependent on the 
$B$-meson parameters, than the individual 
form factors. Our predictions 
are in agreement with the results obtained from the 
LCSR using DA's of strange and nonstrange light mesons; 
e.g., Eq.~(\ref{fplratio}) can be compared 
with $f^+_{B K}(0)/f^+_{B \pi}(0)= 1.36^{+.12}_{-.09}$ 
obtained in \cite{KMM} and  Eq.~(\ref{Tratio})   
with the most recent result \cite{BZnew}:
$T_1^{B K^*}(0)/T_1^{B \rho}(0)=1.17\pm 0.09$\,.
These are very important checks, because 
the new sum rules are independent of the $SU(3)$-violating 
Gegenbauer moments of the kaon and $K^*$.

In addition, returning to the $B\to \pi$ transition
we predict the combination
\be 
\frac{f_B}{f_{B\pi}^+(0)\lambda_B}= 1.56 \pm 0.17\,,
\label{eq-combi}
\ee 
which determines
the coefficient of the hard-scattering contribution 
to the $B\to\pi\pi$ amplitude in the QCD factorization 
approach (for a recent analysis, see e.g., \cite{BJ}). 
Note, that within our method this ratio is practically 
independent of $\lambda_B$ and $f_B$ and is to a 
large extent determined 
by the parameters of the pion channel.

Summarizing, in this paper we have obtained a set of new 
QCD sum rules relating various 
$B\to P,V$ transition form factors
to the universal light-cone DA's of B-meson.
The contributions of the three-particle DA's to the new LCSR  
have been calculated. In addition, we studied the $B$-meson
three-particle DA's, employing QCD sum rules in HQET,
and have obtained a realistic  exponential model of these DA's.

The correlation functions  with an on-shell $B$ meson
and a light-quark current   
allow many other applications to the heavy-light transitions, 
by simply changing the quantum numbers of the light-quark 
current. One does not need to install different light-meson 
DA's, and the two-point sum rules in 
the light-meson channels provide necessary information on 
the duality thresholds. 
With the interval of the inverse moment $\lambda_B$ 
from the QCD sum rules in HQET \cite{GN,BIK}, 
the numerical results obtained in this paper, 
including the $SU(3)$-violating
ratios, provide a nontrivial check 
of the new method with respect to the light-meson LCSR. 

The new $B$-meson LCSR deserve further development.
In this paper only the leading, zeroth order in $\alpha_s$ 
of the light-cone OPE has been taken into account. 
To complete the LCSR derivation at the NLO level, one has  
to calculate the QCD radiative corrections to the correlation function, involving the renormalization effects.
In addition, further light-cone expansion 
of the two- and three-particle heavy-light matrix elements
is desirable, in order to clarify  the role
of yet unaccounted $B$-meson DA's in generating $1/m_b$ and/or 
$1/M^2$ corrections to the sum rules. 
To obtain the necessary elements of these  DA's, 
one can use the technique of HQET sum rules.

\bigskip

\noindent {\bf Acknowledgements}

\noindent We are grateful to 
Vladimir Braun, Thorsten Feldmann and Andrei Grozin 
for useful discussions and comments.
This work was supported by the Deutsche Forschungsgemeinschaft 
under the  contract No. KH205/1-1 and by the German Minister 
of Research (BMBF, contract No. 05HT6PSA).

\section*{Appendix}
Here the expressions for the LCSR at $q^2\neq 0$ and  
$ m_1\equiv m\neq 0 $, are presented ($ m_2=0$) : 
 
\begin{itemize}

\item  $B\to P$  form factors of the 
vector transition current  

\begin{multline}
f^+_{BP}(q^2)=
\frac{\text{fB}m_B}{\text{fP}}
\Bigg\{ 
\int\limits_0^{\text{xi0}(q^2,\text{s0})} d\sigma 
\exp\left(\frac{-s(\sigma,q^2)+m_P^2}{\text{M2}}\right) \\
\times \left[
\frac{\text{barxi}^2
\text{mB}^2}{\text{barxi}^2\text{mB}^2+m^2-q^2}\text{phiBmin}(\sigma m_B
) +
    \left(1-\frac{\text{barxi}^2 \text{mB}^2}{
\text{barxi}^2\text{mB}^2+m^2-q^2}\right) \text{phiBp}(\sigma m_B)
\right.
\\
\left.
+\frac{2 \text{barxi} \left(m^2-q^2\right)
\text{mB}}{\left(\text{barxi}^2
    \text{mB}^2+m^2-q^2\right)^2}\text{PhiBpm}(\sigma m_B) \right] 
+\Delta f^+_{BP}(q^2,s_0,M^2)
\Bigg\} \,, 
\label{eq:fplBP1}
\end{multline}

\begin{multline}
f^+_{BP}(q^2)+f^-_{BP}(q^2)=
\frac{\text{fB}m_B}{ \text{fP}} 
\Bigg\{\int\limits_0^{\text{xi0}(q^2,\text{s0})} \, d\sigma 
\exp\left(\frac{-s(\sigma,q^2)+m_P^2}{\text{M2}}\right) \\
\times\left[\frac{(m-2
\sigma \text{barxi} \text{mB} )\text{mB} 
}{\text{barxi}^2 \text{mB}^2+m^2-q^2}\text{phiBmin}(\sigma m_B)
+
\left(1\,-\,\frac{ \sigma }{\text{barxi}}-
\frac{ (m-2 \sigma\text{barxi} \text{mB} )\text{mB} }{\text{barxi}^2
  \text{mB}^2+m^2-q^2}\right) \text{phiBp}(\sigma m_B)
\right.\\
\left.
-2 m_B\left(\frac{\text{barxi} (m-2 
\sigma \text{barxi} \text{mB} )m_B 
    }{\left(\text{barxi}^2 \text{mB}^2+m^2-q^2\right)^2}+\frac{ (
    \sigma -\text{barxi}) }{\text{barxi}^2 \text{mB}^2+m^2-q^2}\right)
    \text{PhiBpm}(\sigma m_B)\right]\\
    +\Delta f^{\pm}_{BP}(q^2,s_0,M^2)
\Bigg \} \,, 
\label{eq:fpmBP1}
\end{multline}

\item $B\to P$ form factor of the tensor current  
\begin{multline}
f^T_{BP}(q^2)=
\frac{\text{fB}(m_B+m_P)m_B^2}{ \text{fP}(
(m_B^2-m_P^2)-q^2)} 
\Bigg\{\int\limits_0^{\text{xi0}(q^2,\text{s0})} \! d\sigma  
\exp\left(\frac{-s(\sigma,q^2)+m_P^2}{\text{M2}}\right) \\
\times   \left[\dfrac{
\text{barxi}^2m_B^2-m^2+(\sigma-\text{barxi})q^2}
{\text{barxi}^2 m_B^2+m^2-q^2}\left(\text{phiBmin}(\sigma m_B)-
\text{phiBp}(\sigma m_B)\right)
\right.\\
\left.
+\dfrac{1}{m_B}\left(2\dfrac{m^2(2 
\text{barxi}m_B^2-q^2)
+q^2(q^2-\text{barxi}(1+\sigma)m_B^2
)
}{(\text{barxi}^2 \text{mB}^2+m^2-q^2)^2}-\dfrac{1}{\text{barxi}}\right)
    \text{PhiBpm}(\sigma m_B)\right]\\
+\Delta f^{T}_{BP}(q^2,s_0,M^2)
\Bigg \} \,,  
\label{eq:fTBP}
\end{multline}

\item $B\to V$ form factor of the vector current  


\begin{multline}
V^{BV}(q^2)=\frac{\text{fB} \text{mB}^2}{2 \text{fV} \text{mV}} 
(\text{mB}+\text{mV})
 \Bigg\{\int\limits_0^{\text{xi0}(q^2,\text{s0})}\, d\sigma  
\exp\left(\frac{-s(\sigma,q^2)+m_V^2}{\text{M2}}\right) \\
\times \left[\frac{m}{\text{barxi}^2\text{mB}^2+m^2-q^2}\text{phiBmin}(\sigma m_B)
+\left(\frac{1}{\text{barxi}m_B}-\frac{ m
    }{\text{barxi}^2 \text{mB}^2+m^2-q^2}\right) \text{phiBp}(\sigma m_B\ )
\right.\\
\left.
-\frac{2 \text{barxi}m\text{mB}}{\left(\text{barxi}^2\text{mB}^2+m^2-q^2\right)^2}
\text{PhiBpm}(\sigma m_B)\right]
+\Delta V^{BV}(q^2,s_0,M^2)\Bigg \} \,,  
\label{eq:VBV}
\end{multline}

\item $B\to V$ form factors of the axial current  
\begin{multline}
A_1^{BV}(q^2)=
\frac{\text{fB} \text{mB}^3}{2 \text{fV} \text{mV}
    (\text{mB}+\text{mV})} 
\Bigg\{\int\limits_0^{\text{xi0}(q^2,\text{s0})}\, d\sigma
\exp\left(\frac{-s(\sigma,q^2)+m_V^2}{\text{M2}}\right) \\
\times
\left[\dfrac{(\text{barxi}m_B+m)^2-q^2}{m_B^2\text{barxi}^2}\left
\{\dfrac{\bar{\sigma} m m_B}{\text{barxi}^2m_B^2+m^2-q^2}\text{PhiBmin}(\sigma m_B\ )
\right.\right.
 \\
\left.\left.
+\left(1-\dfrac{\text{barxi}m m_B}{\text{barxi}^2
    \text{mB}^2+m^2-q^2}\right)\text{PhiBp}(\sigma m_B)\right\}
\right.
 \\
\left.
-4\dfrac{\text{barxi} m^2 m_B}{\left(\text{barxi}^2
\text{mB}^2+m^2-q^2\right)^2}\text{PhiBpm}(\sigma m_B)
\right]
+\Delta A_1^{BV}(q^2,s_0,M^2)
\Bigg \} \,,   
\label{eq:A1BV}
\end{multline}

\begin{multline}
A_2^{BV}(q^2)=\frac{\text{fB}m_B}{2 \text{fV} m_V}(m_B+m_V) 
\Bigg\{\int\limits_0^{\text{xi0}(q^2,\text{s0})} \, d\sigma 
\exp\left(\frac{-s(\sigma,q^2)+m_V^2}{\text{M2}}\right) \\
\times\left[\frac{ (m-2 \text{barxi} \sigma\text{mB})\text{mB} 
}{\text{barxi}^2 \text{mB}^2+m^2-q^2}\text{phiBmin}(\sigma m_B)
\right.\\
\left.
+\left(1\,-\,\frac{\sigma }{\text{barxi}}-
\frac{ (m-2 \text{barxi} \sigma \text{mB})\text{mB} 
}{\text{barxi}^2 \text{mB}^2+m^2-q^2}\right) \text{phiBp}(
\sigma m_B)
\right.
\\
\left.
-2 m_B\left(\frac{\text{barxi} (m-2 \text{barxi}\sigma \text{mB} )m_B
    }{\left(\text{barxi}^2 \text{mB}^2+m^2-q^2\right)^2}+\frac{ (
    \sigma -\text{barxi}) }{\text{barxi}^2 \text{mB}^2+m^2-q^2}\right)
    \text{PhiBpm}(\sigma m_B)\right]
\\
+\Delta A_2^{BV}(q^2,s_0,M^2)
\Bigg \} \,,   
\label{eq:A2BV}
\end{multline}
\item $B\to V$ form factor of the tensor current  
\begin{multline}
T^{BV}_1(q^2)=
\frac{\text{fB}m_B^2}{2 \text{fV}\text{mV}}
\Bigg\{
\int\limits_0^{\text{xi0}(q^2,\text{s0})} \, d\sigma
\exp\left(\frac{-s(\sigma,q^2)+m_V^2}{\text{M2}}\right) \\
\left[\left(1+\dfrac{m}{\text{barxi}m_B}\right)\left\{\dfrac{
\text{barxi}m m_B
}{\text{barxi}^2\text{mB}^2+m^2-q^2}\text{PhiBmin}(\sigma m_B)
\right.\right.
\\
\left.\left.
+\left(1-\dfrac{
\text{barxi}m m_B}{\text{barxi}^2\text{mB}^2+
m^2-q^2}\right)\text{PhiBp}(\sigma m_B)\right\}
\right.\\
\left.
+\dfrac{m}{\text{barxi}^2 \text{mB}^2+m^2-q^2}\left(1-\dfrac{2
\text{barxi}m_B (\text{barxi}m_B+m)}{\text{barxi}^2\text{mB}^2+
m^2-q^2}\right)\text{PhiBpm}(\sigma m_B)
\right]
\\
+\Delta T_1^{BV}(q^2,s_0,M^2)
\Bigg \} \,,  
\label{eq:T1BV}
\end{multline}
\end{itemize}
where, in order to compactify the above expressions we use 
the dimensionless integration variable 
$\sigma=\omega/m_B$ and the following
notations: $\bar{\sigma}=1-\sigma$,  
\be
s(\sigma,q^2)=\sigma\text{mB}^2+\frac{m^2-\sigma q^2 }{\text{barxi}}\,,
\nonumber
\ee

\be
\text{xi0}(q^2,\text{s0})=
\frac{\text{mB}^2-q^2+s_0-\sqrt{4 \left(m^2-s_0\right)
    \text{mB}^2+\left(\text{mB}^2-q^2+s_0\right)^2}}{2 \text{mB}^2}\,,
\nonumber\ee
so that at $m=0$ and $q^2=0$, $s(\sigma,0)=\sigma m_B^2$ and $\text{xi0}(0,\text{s0})=s_0/m_B^2$. 

In the above $\Delta f_{BP}^+$,$\Delta f_{BP}^{\pm}$, $\Delta f_{BP}^T$,
$\Delta V^{BV}$, $\Delta A_1^{BV}$, $\Delta A_2^{BV}$, 
and $\Delta T_1^{BV}$ denote the contribuitons of
the  $B$-meson three-particle DA's.
We obtain the following generic formula for these correction (
$\Delta F=\Delta f_{BP}^+,\Delta f_{BP}^{\pm},$ etc.): 
\begin{multline}
\Delta F(q^2,s_0,M^2)\: =\:\int\limits_0^{\text{xi0}(q^2,s_0)}\, d\sigma 
\exp\left(\frac{-s(\sigma,q^2)+m_{P(V)}^2}{\text{M2}}\right) \\
\times\left( -I^{(F)}_1(\sigma) +\frac{I^{(F)}_2(\sigma)}{M^2}-
\frac{I^{(F)}_3(\sigma)}{2M^4}\right)
\\
+ \frac{e^{(-s_0+m_{P(V)}^2)/M^2}}{m_B^2}\Bigg\{\eta(\sigma)\Bigg[ I_2^{(F)}(\sigma)\\
-\frac12\left( \frac{1}{M^2} 
+\frac{1}{m_B^2}\frac{d\eta(\sigma)}{d\sigma}\right)I_3^{(F)}(\sigma)
-\frac{\eta(\sigma)}{2m_B^2}\frac{dI_3^{(F)}(\sigma)}{d\sigma}\Bigg]\Bigg\}\Bigg 
|_{\sigma=\sigma_0}\,,
\end{multline}
where
\be
\eta(\sigma)= \left(1+\frac{m^2-q^2 }{\bar{\sigma}^2 m_B^2}\right)^{-1}\,,
\ee 
and the integrals
over the three-particle DA's multiplying the inverse powers 
of the Borel parameter $1/M^{2(n-1)}$ with $n=1,2,3$
are defined as:
\begin{multline}
I^{(F)}_n(\sigma)= \frac{1}{\bar{\sigma}^n}\int\limits_0^{\sigma m_B} d\omega 
\int\limits_{\sigma m_B-\omega}^{\infty} 
\frac{d\xi}\xi \Bigg[ C^{(F,\Psi A)}_n(\sigma,u,q^2)\Psi_A^{B}(\omega,\xi)\\
+C^{(F,\Psi V)}_n(\sigma,u,q^2)\Psi_V^{B}(\omega,\xi) 
\\
+C^{(F,XA)}_n(\sigma,u,q^2)\overline{X}_A^{B}(\omega,\xi)
+C^{(F,YA)}_n(\sigma,u,q^2)\overline{Y}_A^{B}(\omega,\xi)
\Bigg]\Bigg|_{u=(\sigma m_B -\omega)/\xi }
\label{intn}
\end{multline}
where:
$$\overline{X}_A^{B}(\omega,\xi)=\int\limits_0^\omega d\tau X_A^{B}(\tau,\xi),~~\overline{Y}_A^{B}(\omega,\xi)=\int\limits_0^\omega d\tau Y_A^{B}(\tau,\xi).$$

The nonvanishing coefficients entering Eq.~(\ref{intn}) are: 
\ba
C_2^{(f^+_{BP},\Psi A)}&=&
 \left(-\frac{2 m}{m_B}+\text{ubar}-2 \text{ubar} u\right)\,,~~
%
C_2^{(f^+_{BP},\Psi V)}=
 \left(\frac{2 m}{m_B}+\text{ubar}\right)\,,
\nonumber \\
C_2^{(f^+_{BP},XA)}&=&
\frac{1-2 u}{m_B}\,,~~
C_3^{(f^+_{BP},XA)}=
2\frac{ \left(m^2+q^2-m_B^2 \text{ubar}^2\right) (2 u-1)}{m_B}\,,
\nonumber \\
C_3^{(f^+_{BP},YA)}&=& 
4  (2 m+m_B \text{ubar} (2 u-1))\text{ubar} \,,
\\
\nonumber\\
C_2^{(f_{BP}^{\pm}, \Psi A)}&=&1-\frac{4 m}{m_B}+\text{ubar} (2-4 u)+2 u
\,,~~
C_2^{( f_{BP}^{\pm}, \Psi V)}=1+\frac{4 m}{m_B}+2 \text{ubar}-4 u
\,,
\nonumber \\
C_2^{(f_{BP}^{\pm},XA)}&=&-\frac{2 (\text{ubar}-1) (2 u-1)}{m_B\text{ubar}}\,,~~
\nonumber \\
C_3^{(f_{BP}^{\pm},XA)}&=&-\frac{2}{m_B\text{ubar}} \Big(-(2 \text{ubar}+1)
  (2 u-1) m^2+2 m m_B\text{ubar} +
\nonumber\\
&&
(2\text{ubar}-1) \left(m_B^2 \text{ubar}^2-q^2\right) (2u-1)\Big)
\nonumber \\
C_3^{(f_{BP}^{\pm},YA)}&=&4 \Big (m (4 \text{ubar}-1)+2 m_B (\text{ubar}-1) \text{ubar}
    (2 u-1)\Big)\,,
\\
\nonumber\\
C_1^{(f_{BP}^{T}, \Psi A)}&=&\frac{2 u}{m_B^2 \text{ubar}}\,,
~~C_2^{(f_{BP}^{T}, \Psi A)}=-\frac{2 \left(-m^2+m_B^2 \text{ubar}^2+q^2 (1-2 \text{ubar})\right)
    u}{m_B^2 \text{ubar}}\,,
\nonumber \\
C_1^{( f_{BP}^{T}, \Psi V)}&=&-\frac{2 u}{m_B^2 \text{ubar}}\,,
~~
C_2^{( f_{BP}^{T}, \Psi V)}=-\frac{2 \left(m^2-m_B^2 \text{ubar}^2+q^2 (2 \text{ubar}-1)\right)
    u}{m_B^2 \text{ubar}}\,,
\nonumber \\
C_2^{(f_{BP}^{T},XA)}&=&-\frac{4 m}{m_B^2\text{ubar}}\,,
~~C_3^{(f_{BP}^{T},XA)}=-\frac{4 m \left(m^2-m_B^2 \text{ubar}^2+q^2 (2
    \text{ubar}-1)\right)}{m_B^2\text{ubar}}\,,
\nonumber \\
C_2^{(f_{BP}^{T},YA)}&=&-4 \frac{ (2 u-1)}{m_B}\,,
\nonumber \\
C_3^{(f_{BP}^{T},YA)}&=&4 \frac{ \left(-m^2+m_B^2 \text{ubar}^2+q^2 (1-2\text{ubar})\right) (2 u-1)}{m_B}\,,
\label{CBPTA}
\ea

\ba
C_2^{( V^{BV},\Psi A)}&=&\frac{2 u-1}{m_B}\,,~~
C_2^{( V^{BV}, \Psi V)}=-\frac{1}{m_B}\,,~~
C_2^{(V^{BV},XA)}=-2\frac{(2u-1)}{m_B^2\text{ubar}}\,,~~
\nonumber \\
C_3^{(V^{BV},XA)}&=& -\frac{2 }{m_B^2\text{ubar}}\Big((2 u-1) m^2-
2 m_B \text{ubar} m+\left(m_B^2
\text{ubar}^2-q^2\right) (2 u-1)\Big)\,,
\nonumber \\
C_3^{(V^{BV},YA)}&=&-4\frac{ m}{ m_B}\,,
\ea

\ba
C_1^{( A_1^{BV},\Psi A)}&=&\frac{2 u-1}{m_B^2\text{ubar}}\,,
\nonumber \\
C_2^{( A_1^{BV},\Psi A)}&=&\frac{m^2(2 u-1) 
+2 m m_B \text{ubar} 
-\left(q^2-m_B^2
    \text{ubar}^2\right) (2 u-1)}{m_B^2\text{ubar}}\,,
\nonumber \\
C_1^{( A_1^{BV}, \Psi V)}&=&-\frac{1}{m_B^2\text{ubar}}\,,
~~~C_2^{( A_1^{BV}, \Psi V)}= -\frac{ \left(m^2+2 m m_B \text{ubar} 
-q^2+m_B^2\text{ubar}^2\right)}{m_B^2\text{ubar}}\,,
\nonumber \\
C_1^{(A_1^{BV},XA)}&=&-2\frac{(2u-1)}{m_B^3\text{ubar}^2}\,,
\nonumber \\
C_2^{(A_1^{BV},XA)}&=&-\frac{2 \left(2 m^2-2 q^2+m_B^2 \text{ubar}^2\right) (2u-1)}{m_B^3\text{ubar}^2}\,,
\nonumber \\
C_3^{(A_1^{BV},XA)}&=& -\frac{2 \left(m^4-2 m^2\left(q^2+
m_B^2 \text{ubar}^2\right)
+\left(q^2-m_B^2 \text{ubar}^2\right)^2\right) (2
    u-1)}{m_B^3\text{ubar}^2}\,,
\nonumber \\
C_2^{(A_1^{BV},YA)}&=& -\frac{4 (m+
m_B \text{ubar} (1-2 u))}{m_B^2\text{ubar}}\,,
\nonumber \\
C_3^{(A_1^{BV},YA)}&=&-\frac{4 m  \left(m^2+2 m_B \text{ubar} (2 u-1)
    m-q^2+m_B^2 \text{ubar}^2\right)}{m_B^2\text{ubar}}\,,
\ea

\ba
C_2^{(A_2^{BV},\Psi A)}&=& -\left(3-\frac{4 m}{m_B}-2 u+\text{ubar} (4 u-2)
\right)\,,
\nonumber \\
C_2^{( A_2^{BV}, \Psi V)}&=& -\left(3+\frac{4 m}{m_B}-2 \text{ubar}-4 u
\right)\,,
\nonumber \\
C_2^{(A_2^{BV},XA)}&=&-\frac{2 (\text{ubar}-1) (2 u-1)}{m_B\text{ubar}}\,,
\nonumber \\
C_3^{(A_2^{BV},XA)}&=& -\frac{2}{m_B\text{ubar}} \big(-(2 \text{ubar}+1) (2 u-1) m^2+
2 m m_B \text{ubar} 
\nonumber\\
&&+(2\text{ubar}-1) \left(m_B^2 \text{ubar}^2-
q^2\Big) (2u-1)\right)\,,
\nonumber \\
C_3^{(A_2^{BV},YA)}&=&4 \Big(m (3-4 \text{ubar})+
2 m_B (\text{ubar}-1) \text{ubar}(2 u-1)\Big)\,,
\ea

\ba
C_2^{(T_1^{BV},\Psi A)}&=&
\frac{ (m+m_B \text{ubar} (2 u-1))}{m_B}\,,~~~
C_2^{( T_1^{BV}, \Psi V)}=
-\frac{ (m+m_B \text{ubar})}{m_B}\,,
\nonumber \\
C_2^{(T_1^{BV},XA)}&=&
\frac{1-2 u}{m_B}-\frac{2 m}{m_B^2\text{ubar}}\,,
\nonumber \\
C_3^{(T_1^{BV},XA)}&=&
-\frac{2}{m_B^2\text{ubar}} \Big(m^3+m^2 m_B \text{ubar} (1-2 u) 
-m\left(q^2+m_B^2
    \text{ubar}^2\right) 
\nonumber\\
    &&+m_B \text{ubar} \left(m_B^2
    \text{ubar}^2-q^2\right) (2 u-1)\Big)\,,
\nonumber \\
C_2^{(T_1^{BV},YA)}&=&
2 \frac{(2 u-1)}{m_B}\,,~~~
C_3^{(T_1^{BV},YA)}=-4 \frac{m  (m_B \text{ubar}+m (2 u-1))}{m_B}\,.
\ea

\end{document}